\UseRawInputEncoding
\documentclass[%
 twocolumn,
nofootinbib,
 amsfonts,amsmath,amssymb,
 aps,
prd,
floatfix,
showpacs,
longbibliography
]{revtex4-2}

\usepackage[T1]{fontenc}
\usepackage{graphicx}
\usepackage{dcolumn}
\usepackage{bm}
\usepackage[dvipsnames]{xcolor}
\definecolor{ReflexBlue}{rgb}{ .0902,.0902,.5882}
\usepackage[breaklinks]{hyperref}
\hypersetup{
    colorlinks=true,
    linkcolor=ReflexBlue,
    filecolor=magenta,      
    urlcolor=ReflexBlue,
    citecolor=ReflexBlue,
}
\begin{document}

\title{Semiclassical instability of inner-extremal regular black holes}

\author{Tyler McMaken}
 \email{tyler.mcmaken@colorado.edu}
\affiliation{%
 JILA and Department of Physics, University of Colorado, Boulder, Colorado 80309, USA
}%
\date{\today}

\begin{abstract}
The construction of black hole spacetimes that are regular (singularity-free) is plagued by the ``mass inflation'' instability, a classical perturbation instability induced by the surface gravity at the inner horizon and characterized by exponentially diverging stress-energy there. Recently, a class of ``inner-extremal'' regular black holes was proposed that possesses a vanishing inner-horizon surface gravity and therefore avoids mass inflation, while still maintaining a horizon separation and a non-zero outer-horizon surface gravity. However, when semiclassical effects are taken into account, it is found that an inner-horizon instability remains for generic inner-extremal regular black holes formed from collapse. This semiclassical divergence is analyzed from the perspective of both the effective Hawking temperature and the renormalized stress-energy tensor, and its origin and genericity are examined in detail.
\end{abstract}

\maketitle

\section{Introduction}
\label{sec:int}
\subsection{Classical instabilities}
\label{subsec:classical}
In 1965, Penrose detailed the conditions under which a black hole must possess within its horizon a spacetime curvature singularity, where the laws of general relativity break down and demand a more complete theory of quantum gravity \cite{pen65}. Subsequently, proposals for so-called ``regular'' black holes attempted to circumvent the conditions of the singularity theorem so that no appeal to higher theories of gravity would be needed. A common path to doing so is the violation of global hyperbolicity through the presence of a Cauchy horizon (which will be subsequently referred to as an ``inner horizon''; the technical distinction between the two terms is irrelevant here). In the case of spherical symmetry, an inner horizon is in fact required of any regular black hole solution \cite{dym02,car20a,car20b}.

The problem with the presence of an inner horizon within a regular black hole, as first pointed out by Penrose just a few years after arriving at his singularity theorem, is that the inner horizon is a surface of infinite blueshift \cite{pen68,sim73}. Any external perturbations to the spacetime will produce ingoing radiation that an outgoing observer approaching the inner horizon will detect with exponentially diverging energy. Subsequent perturbation models from Poisson-Israel \cite{poi90}, Ori \cite{ori91}, and Hamilton \cite{ham10} analyzed different facets of this effect (known as the ``mass inflation'' instability) in more detail, finding that the inner horizon becomes singular whenever its surface gravity is non-zero due to interactions between ingoing and outgoing perturbations.

In order to circumvent the mass inflation problem, a number of regular black hole solutions have been recently developed that possess an inner horizon with zero surface gravity, first in the spherical case \cite{car22} and subsequently in the rotating case \cite{fra22}. For a static, spherically symmetric black hole with line element
\begin{equation}\label{eq:SSS}
    ds^2=-\Delta(r)dt^2+\frac{dr^2}{\Delta(r)}+r^2\left(d\theta^2+\sin^2\!\theta d\varphi^2\right),
\end{equation}
the horizon function ${\Delta(r)}$ contains zeros at the locations of the horizons (at ${r=r_+}$ for the outer horizon and ${r=r_-}$ for the inner horizon) and asymptotes to unity as ${r\to\infty}$ (assuming the spacetime is asymptotically flat). The (generalized) surface gravity $\kappa$ at any radius $r$ in this spacetime is defined by
\begin{equation}\label{eq:surfacegravity}
    \kappa(r)\equiv\frac{1}{2}\frac{d\Delta}{dr},
\end{equation}
so in order for $\kappa$ to vanish at the inner horizon, the horizon function must contain a degenerate root at that horizon. Such a condition is satisfied for extremal black holes, where the inner horizon coincides with the outer horizon (${r_+=r_-}$), but if one wishes to keep the outer horizon sufficiently separated from any exotic quantum gravitational physics modifying the inner horizon (and indeed, neither have extremal black holes been observed in nature nor should be they theoretically possible by the third law of black hole thermodynamics), the next-simplest choice for the horizon function is a triple root at $r_-$:
\begin{equation}\label{eq:Delta_SIERBH}
    \Delta(r)=\frac{(r-r_+)(r-r_-)^3}{F(r)},
\end{equation}
where
    $$F(r)\equiv(r-r_+)(r-r_-)^3+2Mr^3+\left(a_2-3r_-(r_++r_-)\right)r^2$$
\cite{car22}. Here $M$ is the mass of the black hole and $a_2$ is a real parameter that must satisfy
\begin{equation}
    a_2\gtrsim\frac{9}{4}r_+r_-
\end{equation}
in order for the horizon function to contain no poles along the real axis. The authors of Ref.~\cite{car22} additionally assume that $r_+$ lies in the vicinity of $2M$, while $r_-$ lies in the vicinity of 0. With these choices, we thus have an ``inner-extremal'' regular black hole that behaves approximately like Schwarzschild outside the outer horizon but contains a regular de Sitter core within, fine-tuned so that ${\kappa(r_-)=0}$. In particular, near ${r=0}$, the spacetime possesses a cosmological constant
\begin{equation}
    \Lambda=3\ \frac{a_2-3r_-(r_++r_-)}{r_+r_-^3},
\end{equation}
while all remaining stress-energy contributions to the spacetime curvature vanish.

The story for the case of rotating inner-extremal regular black holes \cite{fra22} is similar to the spherical case, except that the authors of Ref.~\cite{fra22} include an additional conformal factor to maintain regularity at ${r=0}$ so that the horizon function can be fine-tuned independently from the additional regularity constraint (more details are provided in Sec.~\ref{subsec:rot}). The conclusion of the matter for both models is that the black holes remain classically stable to perturbations that would otherwise cause mass inflation at the inner horizon. It should also be mentioned that these black holes are marginally stable to the classical kink instability \cite{mae05}, which generally applies to black holes with ${\kappa(r_-)<0}$.

\subsection{Semiclassical instabilities}
\label{subsec:semiclassical}
Despite the classical stability of inner-extremal regular black holes, far more dangerous instabilities present themselves when semiclassical effects are taken into account. The most recognizable semiclassical effect one may wish to include is the evaporation of the black hole due to Hawking radiation from the outer horizon. Such an evaporation has been incorporated into regular black hole models like the Hayward metric in Refs.~\cite{fro17b,bon23} by adding a time dependence to the mass parameter. In these models, the influence of Hawking radiation dominates that of the mass inflation Price tail at asymptotically late times, leading to one of three results: as the outer horizon shrinks to meet the inner horizon, either the black hole will evaporate entirely (the so-called ``sandwich'' model \cite{fro17a}) and leave an unphysically large burst of energy from the inner horizon, or the black hole will form a cold, stable, extremal remnant where mass inflation either vanishes or is tamed to a power law instead of the usual exponential divergence.

However, a first-order mass loss from Hawking evaporation is not the only possible semiclassical effect, and especially close to the inner horizon, back-reactions from quantum fields there play a much more crucial role in the geometry's dynamical evolution. A common approach to analyzing semiclassical perturbations self-consistently is to construct an additional covariant term ${\langle T_{\mu\nu}\rangle^{\text{ren}}}$ contributing to the stress-energy of the Einstein equations,
\begin{equation}\label{eq:semi_einstein}
    G_{\mu\nu}=8\pi\left(T_{\mu\nu}^{\text{class}}+\langle T_{\mu\nu}\rangle^{\text{ren}}\right),
\end{equation}
where ${\langle T_{\mu\nu}\rangle^{\text{ren}}}$ represents the renormalized vacuum expectation value of the stress-energy tensor for some quantum field. The calculation of ${\langle T_{\mu\nu}\rangle^{\text{ren}}}$ is generally not an easy task, but it has been shown numerically that the flux components of ${\langle T_{\mu\nu}\rangle^{\text{ren}}}$ physically diverge at the inner horizon of Reissner-Nordstr{\"o}m \cite{zil20,hol20} and Kerr \cite{his80,zil22b} black holes, leading to an even stronger singularity at the inner horizon than that imposed by mass inflation.

The conclusion that the semiclassical inner horizon instability leads to a strong singularity relies on the assumption that the inner horizon remains sufficiently static in comparison to the timescale at which the divergent semiclassical flux precipitates. What if such a condition is not met when a dynamical back-reaction is included? Classically, dynamical mass inflation tends to push the inner horizon inward until it is close enough to ${r=0}$ and moving slowly enough that a singularity can form (though certain regular black hole models may lead to asymptotically finite internal mass parameters) \cite{bar22,bon21}. But semiclassically, an analysis of the initial tendencies of Eq.~\eqref{eq:semi_einstein} indicate that the inner horizon should evaporate outward to meet the outer horizon on very rapid timescales \cite{bar21,bar22}. If this semiclassical inflation is strong and quick enough to overcome classical inflation and reach equilibrium before higher-order quantum gravity takes over, the perturbed, collapsing body may stabilize into either an extremal black hole or a compact horizonless object.

The question that may now be asked is whether models of regular black holes that are not subject to the classical mass inflation instability will also be stable to semiclassical perturbations. As will be seen throughout the course of this analysis, the answer is a resounding no. Any relevant semiclassical quantity one might evaluate at the inner horizon will contain at least one component that diverges, since such quantities depend not only on the inner horizon's surface gravity, but also on the outer horizon's surface gravity and on the general causal structure of the spacetime. Importantly, it will be found that in a collapse state, any semiclassical, non-extremal black hole model with an inner horizon will feature a divergence at that horizon. The effect of this divergence is that these inner-extremal regular black holes (along with any other classically consistent models) will either evolve to form a singularity at the inner horizon or else will be subject to the same transient effects discussed in Refs.~\cite{bar22}.

The two relevant semiclassical quantities that will be analyzed here are the effective Hawking temperature $\kappa_{\text{eff}}$ and the renormalized stress-energy tensor ${\langle T_{\mu\nu}\rangle^{\text{ren}}}$. Section~\ref{sec:eff} focuses on the analysis of $\kappa_{\text{eff}}$, which tracks the semiclassical effect of particle creation observed at the inner horizon (akin to the Hawking effect observed asymptotically far away), while Sec.~\ref{sec:ren} analyzes the renormalized stress-energy tensor both analytically in the Polyakov approximation and numerically with pragmatic mode-sum renormalization. Finally, the paper concludes in Sec.~\ref{sec:out} with a discussion of the implications and outlook of these calculations.

\section{Effective Hawking temperature}
\label{sec:eff}
Consider the semiclassical effect of particle production, governed by the Bogoliubov coefficients between the modes from a vacuum state and those of an observer, within inner-extremal regular black holes. The calculation of this effect turns out to be feasible enough that it can be performed analytically for an observer at any point in the spacetime, and while it has not been explicitly proven that the perceived radiation will feed back into the geometry's evolution, its Lorentz covariance in the radial case \cite{mcm23} and its effectiveness at reproducing and clarifying known results offer every indication that its effects are genuine, especially in light of its qualitative agreement with the calculations of ${\langle T_{\mu\nu}\rangle^{\text{ren}}}$ in Sec.~\ref{sec:ren} (in fact, ${\langle T_{\mu\nu}\rangle^{\text{ren}}}$ can be directly associated with the effective temperature and its first derivatives \cite{bar16}).

Additionally, note that while the effective temperature does not make use of any response function or renormalization condition, a full calculation for an Unruh-DeWitt detector (linearly coupled to the proper time derivative of a massless scalar field) approaching the inner horizon has been carried out for a general spherically symmetric black hole in 1+1 dimensions \cite{jua22}, with identical conclusions to what is given in Sec.~\ref{subsec:sph}: both the detector's transition rate and observed energy density in the Unruh state always diverge at the right leg of the inner horizon (regardless of the surface gravity at either horizon), while they diverge at the left leg of the inner horizon except in the special case ${\kappa(r_-)=\kappa(r_+)}$ (which can never happen in the proposed inner-extremal regular models).

\subsection{Formalism}
\label{subsec:for}
In what follows, attention will be restricted to the behavior of a quantized Klein-Gordon massless scalar field (a similar analysis can in principle be performed for higher-spin fields). When such a field is placed over a fixed black hole spacetime formed from gravitational collapse, Hawking \cite{haw75} showed that a characteristic exponential peeling relation between incoming modes from past null infinity and outgoing modes at future null infinity implies the detection of particles by an asymptotically distant future observer from an asymptotically distant past vacuum state.

At the heart of Hawking's calculation is the idea that an exponential rate of redshift between two vacuum states connected by null geodesics leads to a Planck-distributed Bogoliubov coefficient probability ${|\beta_{\omega\omega'}|^2}$ for those states. While Hawking only considered observers asymptotically far from the black hole, one may in principle choose any observer at any location in the spacetime and use the vacuum state defined by their local frame of reference. Such a formalism was developed in Refs.~\cite{bar11a,bar11b}, in which an effective temperature function was defined as
\begin{equation}\label{eq:kappa_u}
    \kappa_\text{eff}(u)\equiv-\frac{d}{du}\ln \left(\frac{dU}{du}\right),
\end{equation}
governing the exponential rate of change between an observer's outgoing null coordinate $u$ and the null coordinate $U$ of an emitter used to define the vacuum state, where the function $U(u)$ described the null geodesic connecting the two worldlines parametrized by the coordinates $U$ and $u$. As long as this effective temperature $\kappa_\text{eff}$ is suitably adiabatic,\footnote{Note that even if the adiabatic condition is not satisfied for some non-zero effective temperature, the Bogoliubov coefficients are still expected to yield a non-zero detection of particles; the only difference is that the spectral distribution of produced particles will generally be non-thermal (see, e.g., Ref.~\cite{mcm23}).} via the condition
\begin{equation}
    \epsilon(u)\equiv\frac{1}{\kappa_\text{eff}^2}\left|\frac{d\kappa_\text{eff}}{du}\right|\ll1,
\end{equation}
Hawking's exact Bogoliubov coefficient calculation will fall into place and a thermal spectrum will be detected by the observer at the temperature
\begin{equation}
    T_H(u)=\frac{\kappa_\text{eff}(u)}{2\pi}.
\end{equation}

Two modifications to the above formalism will help to simplify the calculation of particle production and make it possible to calculate for both inner-extremal regular black hole models below. First, instead of beginning with a Minkowski vacuum state at past null infinity and connecting null rays through a dynamical collapse geometry, it is common to consider a stationary metric of an eternal black hole (like the Schwarzschild metric, or in this case, a static, regular black hole) and place boundary conditions at the past horizon to mimic the exponential redshifting of the collapsing body's apparent horizon. Such a choice of boundary conditions is referred to as the (past) Unruh vacuum state \cite{unr76} and consists of modes that are positive-frequency with respect to the timelike Killing vector ${\partial/\partial t}$ at past null infinity and with respect to the canonical affine Killing field ${\partial/\partial U}$ along the past horizon.

Secondly, instead of using null coordinates, since both the observer and emitter can naturally use their proper times $\tau_\text{ob}$ and $\tau_\text{em}$ to label the different null rays they encounter throughout their journey, Eq.~\eqref{eq:kappa_u} can be recast in a more intuitive form:
\begin{equation}\label{eq:kappa_tau}
    \kappa_\text{eff}=-\frac{d}{d\tau_\text{ob}}\ln\left(\frac{\omega_\text{ob}}{\omega_\text{em}}\right),
\end{equation}
where the frequency $\omega$ (with either subscripts ``ob'' for an observer or ``em'' for an emitter, which will be dropped hereafter when either label could apply), defined by
\begin{equation}\label{eq:freq}
    \omega\equiv-k^\mu\dot{x}_\mu,
\end{equation}
is the temporal component of a null particle's coordinate 4-velocity ${k^\mu\equiv dx^\mu/d\lambda}$, measured in the frame of an observer or emitter with coordinate 4-velocity ${\dot{x}^\mu\equiv dx^\mu/d\tau}$. The Unruh state can then be encoded by the proper time of an emitter if that emitter is taken to be in free fall from rest at infinity and is placed at either ${r_\text{em}\to\infty}$ (for ingoing modes) or ${r_\text{em}\to r_+}$ (for outgoing modes). For more details on calculations within this formalism, see, e.g., Ref.~\cite{mcm23}.

In what follows, the above formalism will be applied first to spherical inner-extremal regular black holes \cite{car22} in Sec.~\ref{subsec:sph}, and then to rotating inner-extremal regular black holes \cite{fra22} in Sec.~\ref{subsec:rot}.

\subsection{Spherical regular black holes}
\label{subsec:sph}
For the static, spherically symmetric metric encoded by Eq.~\eqref{eq:SSS}, the frequency $\omega$ of Eq.~\eqref{eq:freq} measured in the frame of an observer ($\equiv\omega_\text{ob}$) or emitter ($\equiv\omega_\text{em}$) with energy $E$, normalized to the frequency $\omega_\infty$ seen at rest at infinity, is
\begin{equation}
    \frac{\omega}{\omega_\infty}=\frac{E\pm\sqrt{E^2-\Delta}}{\Delta},
\end{equation}
where the upper (lower) sign applies to outgoing (ingoing) null rays. The effective temperature $\kappa$ can then be calculated with the help of the chain rule \cite{ham18}:
\begin{align}\label{eq:kappa_chainrule}
    \kappa_\text{eff}&=-\frac{d}{d\tau_\text{ob}}\ln\left(\frac{\omega_\text{ob}}{\omega_\text{em}}\right)\nonumber\\
    &=-\omega_\text{ob}\left(\frac{\dot{r}_\text{ob}}{\omega_\text{ob}}\frac{\partial\ln\omega_\text{ob}}{\partial r_\text{ob}}-\frac{\dot{r}_\text{em}}{\omega_\text{em}}\frac{\partial\ln\omega_\text{em}}{\partial r_\text{em}}\right)\nonumber\\
    &=\mp\frac12\frac{\omega_\text{ob}}{\omega_\infty}\left(\frac{d\Delta_\text{ob}}{dr_\text{ob}}-\frac{d\Delta_\text{em}}{dr_\text{em}}\right).
\end{align}

As mentioned in Sec.~\ref{subsec:for}, for outgoing modes (upper sign), the Unruh emitter must be placed at the outer horizon (${r_\text{em}\to r_+}$), and for ingoing modes (lower sign), the Unruh emitter resides at infinity (${r_\text{em}\to\infty}$). The result is the sensation of two independent effective temperatures corresponding to the outgoing ($\kappa_\text{eff}^\text{hor}$) and ingoing ($\kappa_\text{eff}^\text{sky}$) Hawking modes originating from the past horizon below and the sky above the observer, respectively. These effective temperatures for an inertial observer at radius $r$ looking in a radial direction take on the following forms, consisting of a Doppler factor multiplied by an observer-dependent surface gravity and a state-dependent surface gravity:
\begin{subequations}\label{eq:kappaeff_SIERBH}
\begin{align}\label{eq:kappaeffhor_SIERBH}
    \kappa_\text{eff}^\text{hor}(r)&=\frac{-E-\sqrt{E^2-\Delta(r)}}{\Delta(r)}\ \left(\kappa(r)-\kappa(r_+)\right),\\\label{eq:kappaeffsky_SIERBH}
    \kappa_\text{eff}^\text{sky}(r)&=\frac{E-\sqrt{E^2-\Delta(r)}}{\Delta(r)}\ \kappa(r),
\end{align}
\end{subequations}
where ${\kappa(r)}$ is the generalized surface gravity defined by Eq.~\eqref{eq:surfacegravity}.

For an observer at rest far away from the black hole, if the spacetime is asymptotically flat, the outgoing effective temperature $\kappa_\text{eff}^\text{hor}$ of Eq.~\eqref{eq:kappaeffhor_SIERBH} approaches ${\kappa(r_+)}$, while the ingoing effective temperature $\kappa_\text{eff}^\text{sky}$ of Eq.~\eqref{eq:kappaeffsky_SIERBH} vanishes, as predicted by Hawking. But for an observer near one of the black hole's horizons, Eqs.~\eqref{eq:kappaeff_SIERBH} warrant closer examination.

First, consider the effective temperatures seen at the outer horizon $r_+$. An observer crossing the event horizon must have ${E>0}$, so that in the limit ${\Delta\to0}$, the outgoing and ingoing effective temperatures simplify to
\begin{subequations}\label{eq:kappaeff_SIERBH_r+}
\begin{align}\label{eq:kappaeffhor_SIERBH_r+}
    \lim_{r\to r_+}\kappa_\text{eff}^\text{hor}(r)&=-\frac{E\kappa'(r_+)}{\kappa(r_+)},\\\label{eq:kappaeffsky_SIERBH_r+}
    \lim_{r\to r_+}\kappa_\text{eff}^\text{sky}(r)&=\frac{\kappa(r_+)}{2E},
\end{align}
\end{subequations}
where a prime denotes differentiation with respect to $r$. Eq.~\eqref{eq:kappaeffhor_SIERBH_r+} makes the same assumption as Ref.~\cite{car22} that the surface gravity ${\kappa(r_+)}$ of the spherical inner-extremal regular black hole's outer horizon is non-zero; if on the contrary the outer horizon is degenerate, the outgoing effective temperature $\kappa_{\text{eff}}^{\text{hor}}$ will depend heavily on the choice of how limits are taken: if the collapse occurred far enough into the past that the Unruh emitter's position can be treated as fixed at $r_+$, the outgoing effective temperature $\kappa_\text{eff}^\text{hor}$ will diverge as a power law when the outer horizon is degenerate, but once the observer reaches and passes below $r_+$, the effective temperature will instantaneously drop to zero.

While the outer horizon's ingoing effective temperature seen from the sky above is always positive, the sign of the outer horizon's outgoing effective temperature originating from the past horizon below depends on the radial gradient of the outer horizon's surface gravity. Assuming ${\kappa(r_+)}$ takes on a positive, non-zero value, if the horizon function is concave down at the outer horizon, ${\Delta''(r_+)<0}$, then the effective temperature from the horizon will be positive just like that of the sky. But if ${\Delta''(r_+)>0}$, as occurs for Reissner-Nordstr\"om black holes with a charge-to-mass ratio ${Q/M>\sqrt{8/9}}$ and for the inner-extremal regular black holes of Eq.~\eqref{eq:Delta_SIERBH} with sufficiently large $a_2$, the outgoing effective temperature will become negative. Such a sign change coincides with the change in sign of the radial tidal force at the outer horizon from geodesic deviation \cite{cri16} and is a commonly found semiclassical feature (see, e.g., Ref.~\cite{mcm23} and sources therein).

At the inner horizon, the effective temperatures depend strongly on the sign of the observer's energy\textemdash note that ingoing (${E>0}$) and outgoing (${E<0}$) observers passing through the inner horizon will enter into causally separated sectors of the spacetime. For an ingoing, positive-energy observer passing through the left leg of the inner horizon,
\begin{subequations}\label{eq:kappaeff_SIERBH_r-L}
\begin{align}\label{eq:kappaeffhor_SIERBH_r-L}
    \lim_{r\to r_-,\ E>0}\kappa_\text{eff}^\text{hor}(r)=&\frac{E\ n!}{(r-r_-)^n}\left(\frac{\kappa(r_+)-\kappa(r_-)}{\kappa^{(n-1)}(r_-)}\right)\nonumber\\
    &+\mathcal{O}\left(\frac{1}{(r-r_-)^{n-1}}\right),\\\label{eq:kappaeffsky_SIERBH_r-L}
    \lim_{r\to r_-,\ E>0}\kappa_\text{eff}^\text{sky}(r)=&\frac{\kappa(r_-)}{2E},
\end{align}
\end{subequations}
where $n$ denotes the lowest non-zero order of the Taylor expansion for the horizon function ${\Delta(r)}$ about the inner horizon; if ${\Delta(r)}$ can be expanded close to a horizon $r_\pm$ as
\begin{align}\label{eq:Delta_series}
    \Delta(r)&\approx\Delta'(r_\pm)(r-r_\pm)\nonumber\\
    &+\frac{1}{2}\Delta''(r_\pm)(r-r_\pm)^2\nonumber\\
    &+\frac{1}{6}\Delta^{(3)}(r_\pm)(r-r_\pm)^3+...,
\end{align}
then, e.g., the Reissner-Nordstr\"om inner horizon corresponds to ${n=1}$, while the horizon function of Eq.~\eqref{eq:Delta_SIERBH} corresponds to ${n=3}$, since for that inner-extremal regular black hole, the first derivative ${\Delta'(r_-)=0}$, the second derivative ${\Delta''(r_-)=0}$, but the third derivative
\begin{equation}
    \Delta^{(3)}(r_-)=-\frac{6(r_+-r_-)}{2Mr_-^3+\left(a_2-3r_-(r_++r_-)\right)r_-^2}.
\end{equation}

Conversely, an outgoing, negative-energy observer passing through the right leg of the inner horizon has
\begin{subequations}\label{eq:kappaeff_SIERBH_r-R}
\begin{align}\label{eq:kappaeffhor_SIERBH_r-R}
    \lim_{r\to r_-,\ E<0}\kappa_\text{eff}^\text{hor}(r)&=\frac{\kappa(r_+)-\kappa(r_-)}{2E},\\\label{eq:kappaeffsky_SIERBH_r-R}
    \lim_{r\to r_-,\ E<0}\kappa_\text{eff}^\text{sky}(r)&=\frac{E\ n}{r-r_-}+\mathcal{O}\left((r-r_-)^0\right).
\end{align}
\end{subequations}

Finally, in the special case ${E=0}$, where the observer passes through the central intersection of the ingoing and outgoing portions of the inner horizon, $\kappa_\text{eff}^\text{hor}$ always diverges, while $\kappa_\text{eff}^\text{sky}$ vanishes when ${n>2}$, remains finite when ${n=2}$, and diverges when ${n=1}$.

The conclusion of the above asymptotic forms of the inner horizon effective temperatures is that at least one component of $\kappa_\text{eff}$ will always diverge for any choice of inertial observer at the inner horizon. This occurs even when the inner horizon's surface gravity ${\kappa(r_-)}$ vanishes\textemdash the divergence is a direct result of the Penrose blueshift singularity (the divergence of ${\omega_\text{ob}/\omega_\text{em}}$ for an outgoing observer watching ingoing modes while crossing a horizon with $\Delta\to0$), which does not depend on the surface gravity. For an inertial observer falling in from infinity, even if they reach an inner horizon with zero surface gravity, they will still encounter diverging semiclassical radiation because the surface gravity of the \emph{outer horizon} (which governs the exponential peeling of modes from the initial collapse and can be regarded in some sense as the ``source'' of Hawking radiation) is non-zero.

The semiclassical instability of the inner horizon is thus seen to be an even stronger effect than the classical mass inflation instability, since the effective temperature in the Unruh vacuum from quantum radiation at the inner horizon depends not only on the inner horizon's surface gravity, but also on the outer horizon's surface gravity. Even if ${\kappa(r_-)}$ vanishes, a non-zero ${\kappa(r_+)}$ will prevent an ingoing observer's effective temperature from vanishing at the inner horizon; instead, the observer's modes will become ultraviolet-divergent. The only feasible way to prevent such a divergence for an ingoing observer is to require ${\kappa(r_-)=\kappa(r_+)}$, and a quick parity check shows that this can only occur if both surface gravities are identically zero.

\subsection{Rotating regular black holes}
\label{subsec:rot}
For a rotating inner-extremal regular black hole, the authors of Ref.~\cite{fra22} considered two modifications to the Kerr line element in Boyer-Lindquist \cite{boy67} coordinates: first, a conformal factor is included so that the metric is regular at ${r=0}$, and second, the radial horizon function $\Delta(r)=r^2+a^2-2m(r)r$ is modified from its vacuum Kerr value (${m(r)=M}$) in order to fine-tune the inner horizon's surface gravity. The line element can be written in the same form as the standard Kerr line element \cite{car68} times a conformal factor ${\Psi(r,\theta)}$:
\begin{align}\label{eq:lineelement_RIERBH}
    ds^2=\Psi\bigg(\frac{1}{\Delta}dr^2+d\theta^2&+\frac{\sin^2\!\theta}{\Sigma^2}\left((r^2+a^2)\ d\varphi-a\  dt\right)^2\nonumber\\
    &-\frac{\Delta}{\Sigma^2}\left(a\sin^2\!\theta\ d\varphi-dt\right)^2\bigg),
\end{align}
where the zeros of the function
\begin{equation}
    \Sigma(r,\theta)\equiv r^2+a^2\cos^2\!\theta
\end{equation}
give the location of the Kerr ring singularity, which becomes regularized when the conformal factor
\begin{equation}\label{eq:Psi}
    \Psi(r,\theta)\equiv\Sigma(r,\theta)+\frac{b}{r^{2z}}
\end{equation}
contains positive, non-zero constants $b$ and $z$ such that ${z\geq3/2}$. The horizon function ${\Delta(r)}$ now has dimension $[M]^2$ and in the minimal case contains a degenerate root at the inner horizon:
\begin{equation}\label{eq:Delta_RIERBH}
    \Delta(r)=\frac{(r-r_+)(r-r_-)^3}{F(r)},
\end{equation}
where now
\begin{equation}\label{eq:Delta_RIERBH_F}
    F(r)\equiv r^2+r(2M-r_+-3r_-)+\frac{r_+r_-^3}{a^2}.
\end{equation}
Though the exact positions of the inner and outer horizons will not directly affect the results of the present analysis, for completion's sake, the following forms are assumed in Ref.~\cite{fra22} for the outer and inner horizon radii:
\begin{equation}
    r_+=M+\sqrt{M^2-a^2},\quad r_-=\frac{a^2}{M+(1-e)\sqrt{M^2-a^2}},
\end{equation}
such that the outer horizon radius $r_+$ coincides with its standard Kerr value while the inner horizon radius $r_-$ is modified by the parameter $e$, which must satisfy
\begin{equation}
    -3-\frac{3M}{\sqrt{M^2-a^2}}<e<2
\end{equation}
to maintain regularity. If $e$ is negative, the inner horizon radius will lie below its Kerr value of ${M-\sqrt{M^2-a^2}}$, while if $e$ is positive, the inner horizon radius will lie above its Kerr value.

If a test particle has Killing energy per unit mass $E$, Killing angular momentum along the axis of rotation per unit mass $L$, and Carter constant ${K=Q+(aE-L)^2}$ \cite{car68}, its 4-velocity will take the form
\begin{subequations}
\begin{align}
    \dot{t}&=\frac{1}{\Psi}\left(\frac{r^2+a^2}{\Delta}P_r+aP_\theta\right),\\
    \dot{\varphi}&=\frac{1}{\Psi}\left(\frac{aP_r}{\Delta}+\frac{P_\theta}{\sin^2\!\theta}\right),\\
    \dot{r}^2&=\frac{1}{\Psi^2}\left(P_r^2-\left(K+\left(r^2+\frac{b}{r^{2z}}\right)\delta\right)\Delta\right),\\
    \dot{\theta}^2&=\frac{1}{\Psi^2}\left(K-a^2\cos^2\!\theta\ \delta-\frac{P_\theta^2}{\sin^2\!\theta}\right)
\end{align}
\end{subequations}
\cite{fra22}, where
\begin{subequations}
\begin{align}
    P_r(r)&\equiv\left(r^2+a^2\right)E-aL,\\
    P_\theta(\theta)&\equiv L-aE\sin^2\!\theta,
\end{align}
\end{subequations}
and where ${\delta=1}$ for massive particles while ${\delta=0}$ for massless particles (which will be denoted with scripted constants of motion $\mathcal{E}$, $\mathcal{L}$, $\mathcal{K}$ in contrast to the massive particle's constants $E$, $L$, $K$).

For simplicity, consider an infalling (${\dot{r}<0}$) equatorial (${\theta=\pi/2}$, ${\dot{\theta}=0}$) observer, whose Carter constant must satisfy
\begin{equation}
    K=P_\theta^2=(L-aE)^2.
\end{equation}
Additionally, as a natural generalization from the spherical case, assume the observer is looking at a photon which is purely radial in the zero angular momentum frame (${\mathcal{L}/\mathcal{E}=0}$, ${\mathcal{K}/\mathcal{E}^2=a^2}$). Such a photon will be detected by the observer with a frequency given by Eq.~\eqref{eq:freq}, which when normalized to the frequency ${\omega_\infty=\mathcal{E}}$ seen by an observer at rest at infinity can be written as
\begin{align}
    &\frac{\omega}{\omega_\infty}=\frac{aP_\theta}{\Psi}+\nonumber\\
    &\frac{r^2+a^2}{\Psi}\cdot\frac{P_r\pm\sqrt{\left(P_r^2-\left(K+\Psi\right)\Delta\right)\left(1-\frac{a^2\Delta}{(r^2+a^2)^2}\right)}}{\Delta},
\end{align}
where outgoing (ingoing) null geodesics are given by the upper (lower) sign.

The effective temperature of Eq.~\eqref{eq:kappa_tau} can then be calculated with the same chain rule expansion as in Eq.~(\ref{eq:kappa_chainrule}):
\begin{equation}\label{eq:kappaeff_RIERBH}
    \kappa_\text{eff}=-\omega_\text{ob}\left(\frac{\dot{r}_\text{ob}}{\omega_\text{ob}}\frac{\partial\ln\omega_\text{ob}}{\partial r_\text{ob}}-\frac{\dot{r}_\text{em}}{\omega_\text{em}}\frac{\partial\ln\omega_\text{em}}{\partial r_\text{em}}\right).
\end{equation}
The above form of $\kappa_\text{eff}$ assumes that the photon's impact parameters ${\mathcal{L}/\mathcal{E}}$ and ${\mathcal{K}/\mathcal{E}^2}$ remain constant as the observer moves along their trajectory, which may induce additional non-inertial radiative effects as the observer rotates their field of view, as first discussed in Ref.~\cite{ham18}. However, the presence or absence of such effects will not significantly change the asymptotic behavior of $\kappa_\text{eff}$ as the observer approaches a horizon; nor will the particular choice of the (inertial) observer's conserved angular parameters $L$ and $K$. A more exhaustive analysis of all these effects will be presented elsewhere. For the present study, assume a freely falling zero angular momentum observer (ZAMO), with constants of motion ${L=0}$, ${K=a^2}$, and ${E=1}$ or -1 (ingoing or outgoing, respectively).

A useful intermediate result with the above simplifications (suppressing factors of $\omega_\infty$) is
\begin{align}\label{eq:intermediate}
    \frac{\partial\ln\omega}{\partial r}&=\frac{\omega\mp1/2}{\omega\mp1}\Bigg(\left(1-\frac{a^2\Delta}{(r^2+a^2)^2}\right)^{-1}\nonumber\\
    &\times\left(\frac{4r}{r^2+a^2}-\frac{\Delta'}{\Delta}\right)-\frac{\Psi'}{\Psi}\Bigg),
\end{align}
where primes denote differentiation with respect to $r$ and the upper (lower) sign applies to an ingoing (outgoing) observer with positive (negative) energy $E$.

Just as in the spherical case, the Hawking modes contributing to the effective temperature can be divided into two sectors, the ingoing modes originating from an Unruh emitter at ${r_\text{em}\to\infty}$ in the sky above the observer, and the outgoing modes originating from an Unruh emitter at ${r\to r_+}$ seen at the past horizon below the observer. In the ingoing case ($\kappa_{\text{eff}}^{\text{sky}}$), the subtracted term in Eq.~\eqref{eq:kappaeff_RIERBH} (i.e.\ the limit of Eq.~\eqref{eq:intermediate} as an emitter's position ${r=r_\text{em}}$ asymptotically tends to infinity) vanishes, just as it does for spherically symmetric black holes. In the outgoing case ($\kappa_{\text{eff}}^{\text{hor}}$), the subtracted term in Eq.~\eqref{eq:kappaeff_RIERBH} simplifies to
\begin{equation}
    \lim_{r_\text{em}\to r_+}\frac{\dot{r}_\text{em}}{\omega_\text{em}}\frac{\partial\ln\omega_\text{em}}{\partial r_\text{em}}=\kappa(r_+),
\end{equation}
where ${\kappa(r)}$ is the black hole's generalized surface gravity analogous to Eq.~\eqref{eq:surfacegravity}, which for a rotating black hole with Boyer-Lindquist radius $r$ takes the form
\begin{equation}
    \kappa(r)\equiv\frac{1}{2(r^2+a^2)}\frac{d\Delta}{dr}.
\end{equation}

Though the full expression for the effective temperature $\kappa_\text{eff}$ for an arbitrary observer is too complicated to be presented in a meaningful way here, some useful limits can be shown. As the observer's position is taken asymptotically far from the black hole, the observer's frequency $\omega_\text{ob}$ tends to unity while the first term in the parentheses of Eq.~\eqref{eq:kappaeff_RIERBH} vanishes. As a result,
\begin{subequations}
\begin{align}
    \lim_{r\to\infty}\kappa_\text{eff}^{\text{hor}}(r)&=\kappa(r_+),\\
    \lim_{r\to\infty}\kappa_\text{eff}^{\text{sky}}(r)&=0;
\end{align}
\end{subequations}
i.e., the Hawking effect is exactly reproduced for this particular choice of observer and Unruh emitter. When this observer is taken to the event horizon at ${r=r_+}$, assuming the outer horizon is not degenerate,
\begin{subequations}
\begin{align}
    \lim_{r\to r_+}\kappa_\text{eff}^{\text{hor}}(r)&=-\frac{r_+^2+a^2}{\Psi(r_+)}\cdot\frac{\kappa'(r_+)}{\kappa(r_+)},\\
    \lim_{r\to r_+}\kappa_\text{eff}^{\text{sky}}(r)&=\frac{\kappa(r_+)}{2},
\end{align}
\end{subequations}
in exact analog to the spherical case; compare Eqs.~\eqref{eq:kappaeff_SIERBH_r+}. The conformal factor here is defined as $\Psi(r)\equiv\Psi(r,\pi/2)$ from Eq.~\eqref{eq:Psi}.

The effective temperatures seen at the inner horizon then follow suit. The choice of whether an observer enters the ingoing or outgoing portion of the inner horizon depends on the sign of the Hamilton-Jacobi parameter $P_r$, which for a ZAMO is equivalent to the sign of the observer's energy $E$. For an observer with positive energy, with the horizon function $\Delta$ and quadratic function $F$ from Eqs.~\eqref{eq:Delta_RIERBH} and \eqref{eq:Delta_RIERBH_F}, respectively, the inner horizon effective temperatures are
\begin{subequations}\label{eq:kappaeff_RIERBH_r-L}
\begin{align}\label{eq:kappaeffhor_RIERBH_r-L}
    &\lim_{r\to r_-,\ E>0}\kappa_\text{eff}^\text{hor}(r)=-\frac{r_-^2+a^2}{\Psi(r_-)}\nonumber\\
    &\qquad\times\frac{F(r_-)(r_-^2+a^2)(r_+-r_-)^2}{F(r_+)(r_+^2+a^2)(r-r_-)^3}+\mathcal{O}\left(\frac{1}{(r-r_-)^2}\right),\\\label{eq:kappaeffsky_RIERBH_r-L}
    &\lim_{r\to r_-,\ E>0}\kappa_\text{eff}^\text{sky}(r)=0,
\end{align}
\end{subequations}
while for an observer with negative energy, the inner horizon effective temperatures are
\begin{subequations}\label{eq:kappaeff_RIERBH_r-R}
\begin{align}\label{eq:kappaeffhor_RIERBH_r-R}
    \lim_{r\to r_-,\ E<0}\kappa_\text{eff}^\text{hor}(r)=&-\frac{(r_+-r_-)^3}{4(r_+^2+a^2)F(r_+)},\\\label{eq:kappaeffsky_RIERBH_r-R}
    \lim_{r\to r_-,\ E<0}\kappa_\text{eff}^\text{sky}(r)=&-\frac{r_-^2+a^2}{\Psi(r_-)}\cdot\frac{3}{r-r_-}\nonumber\\
    &+\mathcal{O}\left((r-r_-)^0\right).
\end{align}
\end{subequations}
Thus, an inertial, zero angular momentum observer approaching the classically stable inner horizon of a rotating regular black hole will experience a diverging, negative effective Hawking temperature in at least one direction, just as in the spherical case. If the observer is ingoing, the divergence will be seen from the past horizon below them, and if the observer is outgoing, the divergence will be seen from the sky above them.

One may wonder about the generality of these results when different choices for observers and photon trajectories are used, especially since Eq.~\eqref{eq:kappaeff_RIERBH} does not guarantee the constant phase condition that usually warrants a numerical treatment as in Refs.~\cite{ham18,mcm23}. But as it turns out, it can be proven that regardless of the choice of observer or emitter, if the effective temperature seen at the outer horizon is finite, then the effective temperature seen at the inner horizon must diverge somewhere in the observer's field of view. To see why this is the case, a sketch of the proof is given below for an ingoing observer with positive Hamilton-Jacobi parameter $P_r$ (a similar argument can be made for an outgoing observer, \emph{mutatis mutandi}).

The effective temperature $\kappa_{\text{eff}}$ can be written in the form
\begin{equation}\label{eq:kappa_chainrule_gen}
    \kappa_{\text{eff}}=-\omega_\text{ob}\left(\frac{\dot{\omega}_\text{ob}}{\omega_\text{ob}^2}-\frac{\dot{\omega}_\text{em}}{\omega_\text{em}^2}\right),
\end{equation}
where an overdot denotes differentiation with respect to proper time; compare\ Eq.~\eqref{eq:kappa_chainrule}. The precise assumptions about the differentiation (e.g.\ keeping the emitter's affine distance or the observer's viewing angles on the sky fixed) can be left arbitrary. There may in general be extra terms in the parentheses of Eq.~\eqref{eq:kappa_chainrule_gen} that nontrivially couple the observer's and emitter's motions, but one may assume that such terms (e.g.\  ones involving derivatives of the emitted photon's impact parameters with respect to the observer's position) can always be chosen to vanish or cancel out by a suitable choice of viewing direction in the observer's sky (e.g.\ in the spherical case this choice is radially inwards or outwards). The remaining terms in Eq.~\eqref{eq:kappa_chainrule_gen} will then be separable in the observer's and emitter's coordinates.

For an Unruh emitter sending outgoing modes from the outer horizon to the observer, assume that the effective temperature in the direction the observer is looking will be finite when the observer reaches the outer horizon:
\begin{equation}\label{eq:finite_kappa_hor}
    \lim_{r_\text{ob}\to r_+}\kappa_\text{eff}=\mathcal{O}\left(\Delta(r_\text{ob})^0\right).
\end{equation}

The key assumption one must make is that the observer's frequency $\omega_\text{ob}$ for outgoing modes classically diverges at either horizon when normalized to the rest frequency at infinity. At the inner horizon, such a divergence manifests as the Penrose blueshift singularity \cite{pen68,sim73}, while at the outer horizon, the emitter's modes will be infinitely redshifted with respect to the observer. In both cases, the effect can be attributed to the fact that the observer can pass through a horizon in finite proper time while an emitter's tortoise coordinate becomes infinite, which is a feature of any black hole spacetime regardless of the surface gravities at the horizons. The divergence of $\omega_\text{ob}$, governed by the timelike component of the line element, asymptotically behaves as $\Delta(r_\text{ob})^{-1}$.

Thus, if the frequencies of Eq.~\eqref{eq:kappa_chainrule_gen} are expressed as ratios to the rest frequency at infinity, then Eqs.~\eqref{eq:kappa_chainrule_gen} and \eqref{eq:finite_kappa_hor} imply that
\begin{equation}\label{eq:kappar+_asymptotic}
    \lim_{r_\text{ob}\to r_+}\frac{\dot{\omega}_\text{ob}}{\omega_\text{ob}^2}=\frac{\dot{\omega}_\text{em}}{\omega_\text{em}^2}+\mathcal{O}\left(\Delta(r_\text{ob})\right).
\end{equation}
Now, if the observer is taken to the inner horizon, the normalized frequency $\omega_\text{ob}$ will still diverge as $\Delta(r_\text{ob})^{-1}$, and the emitter's contribution to the effective temperature will remain unchanged. Substituting the emitter's contribution to the effective temperature from Eq.~\eqref{eq:kappar+_asymptotic} back into Eq.~\eqref{eq:kappa_chainrule_gen} then reveals that the effective temperature at the inner horizon will always diverge unless the value of $\dot{\omega}_\text{ob}/\omega_\text{ob}^2$ for an infalling observer at the outer horizon is the same as that of the inner horizon:
\begin{equation}
    \lim_{r_\text{ob}\to r_-}\kappa_\text{eff}=-\omega_\text{ob}(r_-)\left(\lim_{r_\text{ob}\to r_-}\frac{\dot{\omega}_\text{ob}}{\omega_\text{ob}^2}-\lim_{r_\text{ob}\to r_+}\frac{\dot{\omega}_\text{ob}}{\omega_\text{ob}^2}\right),
\end{equation}
since $\omega_\text{ob}(r_-)$ is of order ${\Theta(\Delta(r_\text{ob})^{-1})}$. For both spherical and rotating inner-extremal regular black holes, the term $\dot{\omega}_\text{ob}/\omega_\text{ob}^2$ corresponds precisely to the black hole's surface gravity at each horizon, and this quantity is assumed to be non-zero at the outer horizon. As argued for the spherical case, the only way for these quantities to be equal at the outer and inner horizons is if the black hole is extremal, so that the outer horizon is degenerate and both surface gravity terms vanish. But more generally, the sign of $\dot{\omega}_\text{ob}/\omega_\text{ob}^2$ at the outer horizon will always be opposite to the sign of $\dot{\omega}_\text{ob}/\omega_\text{ob}^2$ at the inner horizon\textemdash since the observer's normalized frequency at the outer horizon diverges as ${\Delta(r_\text{ob})^{-1}}$ (which is positive as the infaller approaches $r_+$ and, more importantly, has a positive slope), the rate of change of this frequency with respect to the observer's proper time will also be positive at the outer horizon. But at the inner horizon, ${\Delta(r_\text{ob})}$ is negative and further has a negative slope, so that the rate of change of the frequency will always be negative. Thus, the only way that $\dot{\omega}_\text{ob}/\omega_\text{ob}^2$ will match at both the outer and inner horizons is if it identically vanishes at both hypersurfaces, which necessarily assumes that both horizons are degenerate.

\section{Renormalized stress-energy tensor}
\label{sec:ren}
Although the results of Sec.~\ref{sec:eff} give clear evidence for the inevitability of divergent semiclassical behavior at the inner horizon of inner-extremal regular black holes, one may gain further intuition and confirmation by analyzing the behavior of the vacuum expectation value of the renormalized stress-energy tensor ${\langle T_{\mu\nu}\rangle^{\text{ren}}}$. This quantity is not only free of assumptions about adiabaticity and eikonality, but it is also more directly tied to the effects of quantum back-reaction on the underlying spacetime geometry (via the semiclassical Einstein field Eq.~\eqref{eq:semi_einstein}) and therefore is better suited to addressing the question of black hole stability in the semiclassical regime.

As mentioned in Sec.~\ref{sec:int}, the analytic calculation of ${\langle T_{\mu\nu}\rangle^{\text{ren}}}$ is difficult if not impossible for a general spacetime, unless that spacetime possesses a high degree of symmetry. The focus of this analysis will therefore be placed on the evaluation of ${\langle T_{\mu\nu}\rangle^{\text{ren}}}$ for spherical inner-extremal regular black holes, with every expectation (motivated by the results of Sec.~\ref{sec:eff}) that the same tendencies will also be seen in the rotating case.

As a primer, consider the trace anomaly, which helped form the foundations of semiclassical gravity in the early days of quantum field theory in curved spacetimes \cite{duf94}. While the trace of the stress-energy tensor for a classical field with conformal invariance must vanish, the trace of the expectation value of the renormalized stress-energy tensor for a quantum theory with an ultraviolet regulator is generically non-zero\textemdash for a conformal field in four spacetime dimensions, this trace anomaly can be written as
\begin{equation}\label{eq:traceanomaly}
    \langle T^\mu_\mu\rangle=\alpha_F F+\alpha_E E+\alpha_R\mbox{\large$\square$}R
\end{equation}
\cite{duf94}, where $F$ is the squared Weyl tensor, $E$ is the squared Riemann dual tensor (known as the Euler density), and $\mbox{\large$\square$}R$ is the d'Alembertian of the Ricci scalar $R$. These quantities can be expressed in terms of the Riemann tensor $R_{\mu\nu\rho\sigma}$ and the Ricci tensor $R_{\mu\nu}$ as
\begin{subequations}
\begin{align}
    F&=R_{\mu\nu\rho\sigma}R^{\mu\nu\rho\sigma}-2R_{\mu\nu}R^{\mu\nu}+\frac{1}{3}R^2,\\ E&=R_{\mu\nu\rho\sigma}R^{\mu\nu\rho\sigma}-4R_{\mu\nu}R^{\mu\nu}+R^2.
\end{align}
\end{subequations}
The coefficients $\alpha_F$, $\alpha_E$, and $\alpha_R$ depend only on the number of fields and their spins, so that the entire trace anomaly is independent of the vacuum state in which the renormalized stress-energy tensor is evaluated. The form of Eq.~\eqref{eq:traceanomaly} may also contain additional additive terms if the massless fields are coupled to additional background gauge fields.

For a Reissner-Nordstr{\"o}m black hole, the Ricci scalar and its d'Alembertian vanish everywhere, but the squared Weyl tensor and Euler density remain non-zero, so that at the inner horizon, the trace anomaly becomes
\begin{equation}\label{eq:traceanomalyRN}
    \langle T^\mu_\mu\rangle^{\text{RN}}(r_-)=(\alpha_F+\alpha_E)\frac{12(r_+-r_-)^2}{r_-^6}-\alpha_E\frac{8r_+^2}{r_-^6}.
\end{equation}
For a spherical inner-extremal regular black hole, while the Ricci scalar does not vanish (at the inner horizon, ${R=2/r_-^2}$), both $\mbox{\large$\square$}R$ and $E$ do vanish at the inner horizon, so that the trace anomaly simplifies to
\begin{equation}
    \langle T^\mu_\mu\rangle^{\text{IE}}(r_-)=\alpha_F\ \frac{4}{3r_-^4}.
\end{equation}
Note that a finite, non-zero conformal anomaly does not necessarily imply that individual components of a physically realizable renormalized stress-energy tensor will remain well-behaved\textemdash for example, for a Reissner-Nordstr\"om black hole, though ${\langle T^\mu_\mu\rangle}$ from Eq.~\eqref{eq:traceanomalyRN} is finite and non-zero at the inner horizon, the flux components (as well as the trace) of ${\langle T_{\mu\nu}\rangle^{\text{ren}}}$ are well-known to exhibit an inner horizon divergence when a physically realistic vacuum state is used in place of the conformal vacuum \cite{zil20,hol20}.

In principle, one may use the trace anomaly to derive an effective action for a set of auxiliary fields that can be used to define the full covariantly conserved stress-energy tensor ${\langle T_{\mu\nu}\rangle^{\text{ren}}}$ \cite{and07}. However, since inner-extremal regular black holes are not Ricci-flat, the resulting fourth-order differential equations to define ${\langle T_{\mu\nu}\rangle^{\text{ren}}}$ this way do not have analytic solutions in closed form. Further, if the quantum field $\phi$ over the spacetime is not conformally invariant, an additional $\mbox{\large$\square$}\langle\phi^2\rangle^{\text{ren}}$ term must be included in the calculation of the renormalized stress-energy tensor's trace \cite{sel18}. Thus, instead, the renormalized stress-energy tensor will be evaluated two different ways here: first, integrating over the angular degrees of freedom allows for ${\langle T_{\mu\nu}\rangle^{\text{ren}}}$ to be calculated exactly in 1+1 dimensions via the so-called Polyakov approximation (Sec.~\ref{subsec:pol}), and secondly, a pragmatic mode-sum analysis allows for ${\langle T_{\mu\nu}\rangle^{\text{ren}}}$ to be calculated numerically at the inner horizon in the full 3+1 dimensions (Sec.~\ref{subsec:pmr}).

\subsection{Polyakov approximation}\label{subsec:pol}
If the static, spherically symmetric black hole spacetime described by Eq.~\eqref{eq:SSS} is restricted to the $(t,r)$ sector, the stress-energy tensor of a quantized field in the resulting 1+1D spacetime can be uniquely renormalized to yield an exact expression, since the equations of motion for the field are conformally invariant \cite{bar12}. If one converts to a set of double null coordinates $(u,v)$ that define the vacuum state, so that the line element becomes
\begin{equation}\label{eq:metric2D}
    ds^2=-C(u,v)\ du\ dv
\end{equation}
for some conformal factor $C$, the vacuum expectation value of the renormalized stress-energy tensor for a massless, scalar quantum field will be
\begin{subequations}\label{eq:RSET2D}
\begin{align}
    \langle T_{uu}\rangle^{\text{ren}}&=\frac{1}{24\pi}\left(\frac{1}{C}\frac{\partial^2C}{\partial u^2}-\frac{3}{2C^2}\left(\frac{\partial C}{\partial u}\right)^2\right),\\
    \langle T_{vv}\rangle^{\text{ren}}&=\frac{1}{24\pi}\left(\frac{1}{C}\frac{\partial^2C}{\partial v^2}-\frac{3}{2C^2}\left(\frac{\partial C}{\partial v}\right)^2\right),\\
    \langle T_{uv}\rangle^{\text{ren}}&=\frac{1}{24\pi}\left(\frac{1}{C^2}\frac{\partial C}{\partial u}\frac{\partial C}{\partial v}-\frac{1}{C}\frac{\partial^2C}{\partial u\partial v}\right).
\end{align}
\end{subequations}
The contribution made by Polyakov (working in the context of bosonic string theory) was that an effective action for a higher-dimensional theory can be reduced to a two-dimensional, renormalizable, completely integrable theory by performing an averaging sum over all the remaining surfaces \cite{pol81}. In the present context, Polyakov's approximation manifests by averaging over the 2-sphere so that the renormalized stress-energy tensor in 3+1 dimensions is simply given by the expressions of Eqs.~\eqref{eq:RSET2D}, each divided by the factor ${4\pi r^2}$. While such a choice implies that ${\langle T_{\mu\nu}\rangle^{\text{ren}}}$ will behave in a singular fashion at ${r=0}$, this ${r=0}$ singularity at least in the rotating case can only be reached in an infinite proper time \cite{fra22}, but more importantly, it is understood that the renormalized stress-energy tensor in the Polyakov approximation should be further regularized at small $r$ \cite{arr21a}.

\subsubsection{Boulware vacuum}
The calculation of ${\langle T_{\mu\nu}\rangle^{\text{ren}}}$ depends heavily on the choice of vacuum state, which, as mentioned, is dictated by the specification of the conformal factor ${C(u,v)}$ of Eq.~\eqref{eq:metric2D}. One simple choice is to set ${C=\Delta}$ from Eq.~\eqref{eq:Delta_SIERBH}, so that the double null coordinates ${(u,v)}$ coincide with the usual static Eddington-Finkelstein coordinates. The corresponding vacuum state ${|0\rangle_\text{B}}$ is known as the Boulware vacuum, which describes an asymptotically radiation-free black hole as viewed by a static observer in the exterior (and a similar state can be defined for a zero-energy observer in the black hole interior). As a result, the state is not well-defined for an observer at either horizon, and an infaller will see a diverging stress-energy flux at the outer horizon:
\begin{subequations}\label{eq:TmunuB}
\begin{align}\label{eq:TuuB}
    \langle T_{uu}\rangle^{\text{ren}}_\text{B}&=\langle T_{vv}\rangle^{\text{ren}}_\text{B}=\frac{1}{192\pi^2r^2}\left(\kappa'(r)\Delta(r)-\kappa(r)^2\right),\\
    \langle T_{uv}\rangle^{\text{ren}}_\text{B}&=\langle T_{vu}\rangle^{\text{ren}}_\text{B}=\frac{1}{192\pi^2r^2}\kappa'(r)\Delta(r),
\end{align}
\end{subequations}
where $\kappa(r)$ is the generalized surface gravity given by Eq.~\eqref{eq:surfacegravity}. While these null components of ${\langle T_{\mu\nu}\rangle^{\text{ren}}}$ do not diverge at either horizon, the coordinate system does. Changing to a coordinate system that behaves regularly at the horizons, such as the Kruskal-Szekeres coordinates ${(U,V)}$ defined by
\begin{equation}\label{eq:UV}
    \frac{dU}{du}=\text{e}^{-\kappa(r_+)u},\qquad\frac{dV}{dv}=\text{e}^{\kappa(r_+)v},
\end{equation}
reveals that as long as ${\langle T_{uu}\rangle^{\text{ren}}_\text{B}}$ is non-zero at the outer horizon, ${\langle T_{UU}\rangle^{\text{ren}}_\text{B}}$ will diverge as e$^{2\kappa(r_+)u}$ as the horizon at ${u\to\infty}$ is approached. At the outer horizon, the surface gravity ${\kappa(r_+)}$ contributing to Eq.~\eqref{eq:TuuB} remains non-zero, so the Boulware vacuum stress-energy will always diverge in that limit. In accordance with the Fulling-Sweeny-Wald theorem \cite{ful78}, since any Hadamard state should yield finite quantities at the outer horizon, a more astrophysically relevant vacuum state must be sought after.

The two vacuum states that will be used here to find the renormalized stress-energy tensor at the inner horizon are the ``in'' Minkowski vacuum ${|0\rangle_\text{in}}$ and the Unruh vacuum ${|0\rangle_\text{U}}$.

\subsubsection{Minkowski ``in'' vacuum}
The ``in'' vacuum state assumes that asymptotically far into the past, the spacetime is completely flat, with the standard Minkowski vacuum. Then, at a time ${v=v_0}$, an ingoing null shell forms a black hole so that the conformal factor of Eq.~\eqref{eq:metric2D} transitions from ${C(u_\text{in},v_\text{in})=1}$ in the ``in'' region (${v<v_0}$) to ${C(u_\text{out},v_\text{out})=\Delta}$ in the ``out'' region (${v>v_0}$). The corresponding conformal factor of the ``in'' vacuum state to be substituted into Eq.~\eqref{eq:RSET2D} is
\begin{equation}\label{eq:Cin}
    C=\frac{du_\text{out}}{du_\text{in}}\Delta,
\end{equation}
where the relation between the ``in'' and ``out'' coordinates can be found by matching metrics through the collapsing null shell, as detailed below.

The authors of Ref.~\cite{bar21} performed such a matching with sufficient generality by focusing on the asymptotic behavior of ${\langle T_{\mu\nu}\rangle^{\text{ren}}_{\text{in}}}$ at the inner and outer horizons. By expanding the horizon function $\Delta$ about either horizon at $r_\pm$ via the series of Eq.~\eqref{eq:Delta_series}, the stress-energy tensor at $r_\pm$ reduces to
\begin{subequations}\label{eq:Tmunuin}
\begin{align}
    \langle T_{uu}\rangle^{\text{ren}}_{\text{in}}\approx&\ \frac{1}{96\pi^2r_\pm^2}\frac{\kappa''(r_\pm)}{8\kappa(r_\pm)}\left(\text{e}^{2\kappa(r_\pm)(v-v_0)}-1\right)\nonumber\\
    &+\mathcal{O}\left(r-r_\pm\right),\\
    \langle T_{vv}\rangle^{\text{ren}}_{\text{in}}\approx&-\frac{1}{96\pi^2r_\pm^2}\frac{\kappa(r_\pm)^2}{2}+\mathcal{O}\left((r-r_\pm)^2\right),\\
    \langle T_{uv}\rangle^{\text{ren}}_{\text{in}}=\langle T_{vu}\rangle^{\text{ren}}_{\text{in}}\approx&\ \frac{1}{96\pi^2r_\pm^2}\frac{\kappa'(r_\pm)}{2}\text{e}^{\kappa(r_\pm)(v-v_0)}\nonumber\\
    &+\mathcal{O}\left(r-r_\pm\right).
\end{align}
\end{subequations}
For the inner-extremal regular black holes in which ${\kappa(r_-)=0}$, the modified series expansion and subsequent matching procedure lead to the same form for the stress-energy tensor components as that inferred from Eqs.~\eqref{eq:Tmunuin}. In particular, the $uu$-component of the stress-energy tensor at the inner horizon diverges as ${\kappa''(r_-)(v-v_0)}$, while the $vv$-component vanishes. Converting to a set of regular coordinates across the horizon (such as ${\langle T_{rr}\rangle^{\text{ren}}_{\text{in}}}$) yields a similar divergence in $v$. However, as the authors of Ref.~\cite{bar21} note, higher-order terms in the series expansion also contain similar time-dependent divergent factors (except in the expansion of $\langle T_{vv}\rangle^{\text{ren}}_{\text{in}}$), so that the truncated series expansion about the inner horizon becomes less and less of a good approximation as $v$ increases. The opposite happens at the outer horizon, where higher-order time-dependent terms are exponentially damped in accordance with the change in sign of the surface gravity.

To alleviate this problem, instead of performing a series expansion about a general horizon function $\Delta$, consider the exact form of the ``in'' vacuum stress-energy tensor for the specific case of the horizon function of Eq.~\eqref{eq:Delta_SIERBH}. At the null shell boundary, outgoing null geodesics in the ``in'' region satisfy
\begin{equation}
    r=\frac{v_0-u_\text{in}}{2},
\end{equation}
while outgoing null geodesics in the ``out'' region satisfy
\begin{align}\label{eq:tortoise}
    \frac{v_0-u_\text{out}}{2}=&\ r+\frac{A}{(r-r_-)}+\frac{B}{(r-r_-)^2}\nonumber\\
    &+C\ln\left|\frac{r-r_-}{r_-}\right|+D\ln\left|\frac{r-r_+}{r_+}\right|,
\end{align}
where the constants $A$, $B$, $C$, and $D$ define a tortoise coordinate (via ${dr/dr^*=\Delta}$); their exact form in terms of the parameters $r_+$, $r_-$, $a_2$, and $M$ is not too enlightening and will not be given here. After matching these solutions at the null boundary, the resulting stress-energy tensor can then be calculated through Eqs.~\eqref{eq:RSET2D} and \eqref{eq:Cin}. Instead of calculating the full $u$- and $v$-dependence of the conformal factor $C$, one may note that each term on the right-hand side of Eq.~\eqref{eq:Cin} will contribute a separate additive term to the total stress-energy tensor: the contribution from the horizon function $\Delta$ has already been calculated as the static Boulware term of Eqs.~\ref{eq:TmunuB}, and the second state-dependent term will approximately equal the Schwarzian derivative of ${u_\text{in}(u_\text{out})}$, divided by ${-24\pi}$ \cite{fab05}.

The result for the normal stress components (for simplicity the shear stress components are ignored in what follows, since they will generally vanish in the horizon limit) of the renormalized stress-energy tensor in the ``in'' vacuum state, evaluated at the outer horizon (where ${u_\text{in}=v_0-2r_+}$ and ${r=r_+}$), is
\begin{subequations}\label{eq:Tmunuin_rp}
\begin{align}
    \lim_{r\to r_+}\langle T_{uu}\rangle^{\text{ren}}_\text{in}&=0,\\
    \lim_{r\to r_+}\langle T_{vv}\rangle^{\text{ren}}_\text{in}&=-\frac{1}{96\pi^2r_+^2}\frac{\kappa(r_+)^2}{2},
\end{align}
\end{subequations}
while the same components evaluated at the left leg of the inner horizon (where ${u_\text{in}=v_0}$ and ${r=r_-}$) simplify to
\begin{subequations}\label{eq:Tmunuin_rm}
\begin{align}
    \lim_{r\to r_-}\langle T_{uu}\rangle^{\text{ren}}_\text{in}&=\frac{1}{96\pi^2r_-^2}\frac{a_2-3r_-(r_++r_-)}{2r_+r_-^3},\\
    \lim_{r\to r_-}\langle T_{vv}\rangle^{\text{ren}}_\text{in}&=0.
\end{align}
\end{subequations}
The outer horizon value of ${\langle T_{uu}\rangle^{\text{ren}}_\text{in}}$ vanishes because the state-dependent term is proportional to ${\kappa(r_+)^2}$, which exactly cancels the same factor in the Boulware term of Eq.~\eqref{eq:TuuB}, while the inner horizon value of ${\langle T_{vv}\rangle^{\text{ren}}_\text{in}}$ vanishes because both the state-dependent and Boulware terms are identically zero.

At the outer horizon, the interpretation of Eqs.~\eqref{eq:Tmunuin_rp} is that a steady negative ingoing flux counters the outgoing Hawking radiation at infinity and causes the outer horizon to shrink over time, while no outgoing flux is observed at the outer horizon (otherwise, the stress-energy would diverge there when written in coordinates that are regular across the horizon).

At the left leg of the inner horizon, the interpretation of Eqs.~\eqref{eq:Tmunuin_rm} is that the vanishing surface gravity removes any ingoing flux that might shift the position of the inner horizon, but the outgoing flux from the collapse vacuum is non-zero and therefore causes divergent, singular behavior when switching over to Kruskalized coordinates that are regular across the inner horizon.

\subsubsection{Unruh vacuum}
The final vacuum state that will be considered here is the (past) Unruh vacuum ${|0\rangle_\text{U}}$ \cite{unr76}, which is the late-time (${u\to\infty}$) limit of the ``in'' Minkowski state. This state describes the steady-state collapse dynamics of a black hole by replacing the past horizon of an eternal black hole spacetime (such as the inner-extremal regular black hole model) with a semiclassically singular surface that sources exponentially redshifting modes.

The appropriate conformal factor for the Unruh state is
\begin{equation}
    C=\frac{du}{dU}\Delta,
\end{equation}
where $u$ is the standard outgoing Eddington-Finkelstein coordinate and $U$ is the outgoing Kruskal-Szekeres coordinate of Eq.~\eqref{eq:UV}. The resulting components of the renormalized stress-energy tensor are
\begin{subequations}\label{eq:TmunuU}
\begin{align}
    \langle T_{uu}\rangle^{\text{ren}}_\text{U}&=\frac{1}{192\pi^2r^2}\left(\kappa'(r)\Delta(r)-\kappa(r)^2+\kappa(r_+)^2\right),\\
    \langle T_{vv}\rangle^{\text{ren}}_\text{U}&=\frac{1}{192\pi^2r^2}\left(\kappa'(r)\Delta(r)-\kappa(r)^2\right),\\
    \langle T_{uv}\rangle^{\text{ren}}_\text{U}&=\langle T_{vu}\rangle^{\text{ren}}_\text{U}=\frac{1}{192\pi^2r^2}\kappa'(r)\Delta(r).
\end{align}
\end{subequations}
Consider the behavior of Eqs.~\eqref{eq:TmunuU} for the horizon function of Eq.~\eqref{eq:Delta_SIERBH}. At the outer horizon, the only non-zero double-null component of ${\langle T_{\mu\nu}\rangle^{\text{ren}}_\text{U}}$ is the usual ingoing $vv$ term contributing to the shrinking of that horizon. However, at the inner horizon, the only non-vanishing component is the $uu$ component, which is proportional to the square of the \emph{outer} horizon's surface gravity. As a result, conversion to a set of coordinates that are regular across horizons will yield a physical divergence in ${\langle T_{\mu\nu}\rangle^{\text{ren}}_\text{U}}$ along the left leg of the inner horizon. This divergence is of the exact same form as that found in the effective temperature calculations of Eq.~\eqref{eq:kappaeffhor_SIERBH_r-L}\textemdash even though the inner horizon's surface gravity may vanish, the semiclassical flux diverges at the inner horizon because the surface gravity of the outer horizon (which determines the quantum modes' exponential peeling rates) is non-zero.

\subsection{Pragmatic mode-sum renormalization}\label{subsec:pmr}
One may wonder whether the divergence of the renormalized stress-energy tensor at the inner horizon is simply an artifact of the Polyakov restriction to 1+1 dimensions, which does not account for the back-scattering of angular modes. To test whether this is the case, the inner-horizon limit of the 3+1D renormalized stress-energy tensor will be calculated numerically using a prescription developed by Levi and Ori known as pragmatic mode-sum renormalization (PMR) \cite{lev15,lev16b,lev17}.

In the PMR prescription, ${\langle T_{\mu\nu}\rangle}$ is renormalized with covariant point-splitting, where the stress-energy tensor is built out of the field's two-point function and its derivatives. The resulting quantity will formally diverge when the coincidence limit is taken, but it will remain finite when a geometrically constructed counterterm is subtracted from the bare stress-energy tensor. Covariant point-splitting renormalization usually has the numerical difficulty that both the bare stress-energy and the subtracted counterterm formally diverge, so that a finite result can only be obtained when both quantities are regularized to yield analytic closed forms that can be subtracted. The way PMR overcomes this obstacle is by bringing both the bare term and the counterterm under the same mode sum, so that the subtraction can be carried out in a finite fashion mode-by-mode.

If a massless, minimally coupled scalar field $\phi$ is placed over the spherically symmetric spacetime of Eq.~\eqref{eq:SSS} with the inner-extremal regular horizon function of Eq.~\eqref{eq:Delta_SIERBH}, that field will obey the wave equation ${\mbox{\large$\square$}\phi=0}$. Decomposing the field into a sum of modes via
\begin{equation}
    \phi(x)=\sum_{\ell=0}^{\infty}\sum_{m=-\ell}^{\ell}\int_0^\infty\!\!d\omega\ \text{e}^{-i\omega t}Y_{\ell m}(\theta,\varphi)\psi_{\omega\ell}(r)
\end{equation}
leads to the following wave equation for the radial mode functions ${\psi_{\omega\ell}}$:
\begin{equation}\label{eq:wave}
    \frac{d\psi_{\omega\ell}}{dr^{*2}}+\left(\omega^2-\left(\frac{\ell(\ell+1)}{r^2}+\frac{2\kappa}{r}\right)\Delta\right)\psi_{\omega\ell}=0,
\end{equation}
where $r^*$ is the tortoise coordinate defined by ${dr/dr^*=\Delta}$ as in Eq.~\eqref{eq:tortoise}, and ${\kappa(r)}$ is the generalized surface gravity of Eq.~\eqref{eq:surfacegravity}.

The Unruh state for this field $\phi$ is specified by the following boundary conditions on the set of ingoing modes ${\phi_{\omega\ell}^{\text{in}}\equiv\text{e}^{-i\omega t}\psi_{\omega\ell}^{\text{in}}}$ and outgoing modes ${\phi_{\omega\ell}^{\text{up}}\equiv\text{e}^{-i\omega t}\psi_{\omega\ell}^{\text{up}}}$:
\begin{subequations}
\begin{align}
    \phi_{\omega\ell}^{\text{in}}&\to
    \begin{cases}
        0,&\text{past null infinity}\\
        \text{e}^{-i\omega U},&\text{past horizon}
    \end{cases},\\
    \phi_{\omega\ell}^{\text{up}}&\to
    \begin{cases}
        \text{e}^{-i\omega v},&\text{past null infinity}\\
        0,&\text{past horizon}
    \end{cases},
\end{align}
\end{subequations}
with the Kruskal-Szekeres coordinate $U$ of Eq.~\eqref{eq:UV}, the Eddington-Finkelstein coordinate ${u\equiv t-r^*}$ (both in the interior and the exterior), and where the ``past horizon'' denotes the surface for which ${r^*\to-\infty}$ and ${t\to-\infty}$ (both in the interior and the exterior).

In this vacuum state, renormalization of the stress-energy tensor by $\theta$-splitting PMR yields the following formulas for the normal stress components evaluated at the inner horizon:
\begin{subequations}\label{eq:TU_PMR}
\begin{align}\label{eq:TuuU}
    \langle T_{uu}\rangle^{\text{ren}}_{\text{U}}(r_-)&=\sum_{\ell=0}^{\infty}\frac{2\ell+1}{8\pi}\left(\int_0^\infty\!\!d\omega\ \hat{E}^{\text{U}}_{\omega\ell}-\beta\right),\\\label{eq:TvvU}
    \langle T_{vv}\rangle^{\text{ren}}_{\text{U}}(r_-)&=\langle T_{uu}\rangle^{\text{ren}}_{\text{U}}(r_-)-\frac{1}{4\pi r_-^2}\sum_{\ell=0}^{\infty}L_{\ell}^{\text{U}}
\end{align}
\end{subequations}
\cite{zil20}, where 
\begin{subequations}
\begin{align}
    \hat{E}^{\text{U}}_{\omega\ell}=&\ \frac{\omega}{4\pi r_-^2}\Big(|A_{\omega\ell}|^2\left(1+\left(\coth\widetilde{\omega}-1\right)|\rho^{\text{up}}_{\omega\ell}|^2\right)\nonumber\\
    &+\text{csch}\widetilde{\omega}\ \mathfrak{Re}\left(\rho^{\text{up}}_{\omega\ell}A_{\omega\ell}B_{\omega\ell}\right)\Big),\\\label{eq:hawkingoutflux}
    L_{\ell}^{\text{U}}=&\ \frac{2\ell+1}{4\pi}\int_0^\infty\!\!d\omega\ \omega\left(\coth\widetilde{\omega}-1\right)|\tau^\text{up}_{\omega\ell}|^2,
\end{align}
\end{subequations}
where ${\widetilde{\omega}\equiv\pi\omega/\kappa(r_+)}$, and where $\rho^{\text{up}}_{\omega\ell}$, $\tau^\text{up}_{\omega\ell}$, $A_{\omega\ell}$, and $B_{\omega\ell}$ are scattering coefficients described in more detail below.

The above expressions for the components of the renormalized stress-energy tensor at the inner horizon were originally derived for Reissner-Nordstr\"om black holes, but the derivation was carried out with sufficient generality so that it also can be applied to the present case of spherical inner-extremal regular black holes with minimal changes. The most noticeable difference aside from the alternative specification of the horizon function $\Delta$ is in the form of the blind-spot counterterm $\beta$ in Eq.~\eqref{eq:TuuU}, which represents the asymptotic large-$\ell$ plateau value of the integral immediately preceding it. In Reissner-Nordstr\"om, one has ${\beta=\left(\kappa(r_-)^2-\kappa(r_+)^2\right)/(24\pi r_-^2)}$ \cite{zil20}, but the derivation of this analytic expression (in particular, the large-$\ell$ forms of the scattering coefficients derived in Ref.~\cite{sel18}) relies on the Reissner-Nordstr\"om form of the horizon function in several crucial ways. When the inner-extremal horizon function of Eq.~\eqref{eq:Delta_SIERBH} instead is used in the radial wave Eq.~\eqref{eq:wave}, the relevant asymptotic solutions can no longer be written in terms of Bessel functions near the inner horizon (nor any other well-understood special functions). An analytic form for $\beta$ may still be possible for the inner-extremal case through a form of Frobenius matching; however, here it suffices to compute $\beta$ numerically, since the sum of Eq.~\eqref{eq:TuuU} quickly reaches a plateau value within the desired precision after only a few of the lowest-$\ell$ terms are included. Regardless, as will be seen, the divergence of at least one component of ${\langle T_{\mu\nu}\rangle^{\text{ren}}}$ can be shown without making any assumptions about $\beta$.

The scattering coefficients $\rho^{\text{up}}_{\omega\ell}$, $\tau^\text{up}_{\omega\ell}$, $A_{\omega\ell}$, and $B_{\omega\ell}$ are computed by numerically integrating the radial wave Eq.~\eqref{eq:wave} for a set of Eddington-Finkelstein modes propagating between the asymptotic boundaries for both the exterior and interior black hole sectors. In the exterior sector, the reflection coefficient $\rho^{\text{up}}_{\omega\ell}$ gives the fraction of outgoing waves emitted from the outer horizon in the asymptotic past that reflect back to the outer horizon, while the transmission coefficient $\tau^\text{up}_{\omega\ell}$ gives the remaining portion of waves that reach infinity:
\begin{equation}
    \psi_{\omega\ell}^{\text{up}}\to
    \begin{cases}
        \text{e}^{i\omega r^*}+\rho^{\text{up}}_{\omega\ell}\ \text{e}^{-i\omega r^*},&r^*\to-\infty\\
        \tau^{\text{up}}_{\omega\ell}\ \text{e}^{i\omega r^*},&r^*\to\infty
    \end{cases}.
\end{equation}
The reflection coefficient $\rho^{\text{up}}_{\omega\ell}$ is related to the transmission coefficient $\tau^{\text{up}}_{\omega\ell}$ through the condition ${|\rho^{\text{up}}_{\omega\ell}|^2+|\tau^{\text{up}}_{\omega\ell}|^2=1}$ resulting from Wronskian conservation of solutions for the radial wave Eq.~\eqref{eq:wave}.

In the interior sector, where $r^*$ becomes a timelike coordinate, free incoming waves at the outer horizon scatter into a superposition of ingoing and outgoing waves at the inner horizon with the corresponding reflection and transmission coefficients $A_{\omega\ell}$ and $B_{\omega\ell}$:
\begin{equation}
    \psi_{\omega\ell}^{\text{up}}\to
    \begin{cases}
        \text{e}^{-i\omega r^*},&r^*\to-\infty\\
        A_{\omega\ell}\ \text{e}^{i\omega r^*}+B_{\omega\ell}\ \text{e}^{-i\omega r^*},&r^*\to\infty
    \end{cases}.
\end{equation}
For these interior scattering coefficients, which need not remain bounded, the Wronskian condition implies that ${|B_{\omega\ell}|^2-|A_{\omega\ell}|^2=1}$.

\begin{figure*}[t]
\begin{centering}
\includegraphics[width=.45\textwidth]{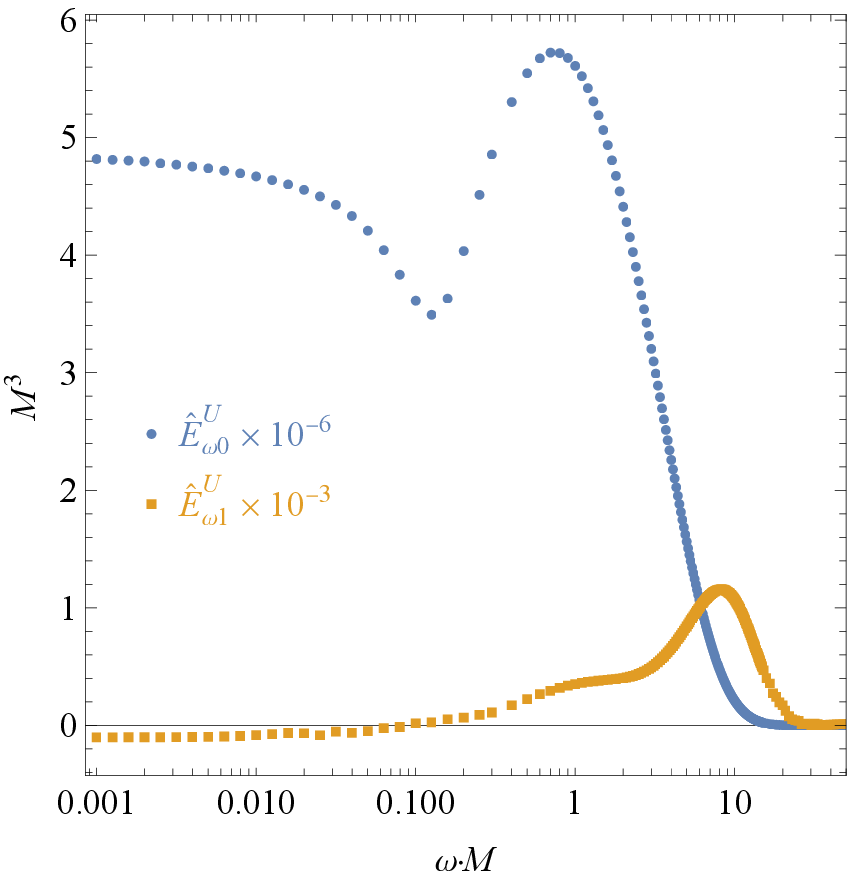}
\hfill
\includegraphics[width=.49\textwidth]{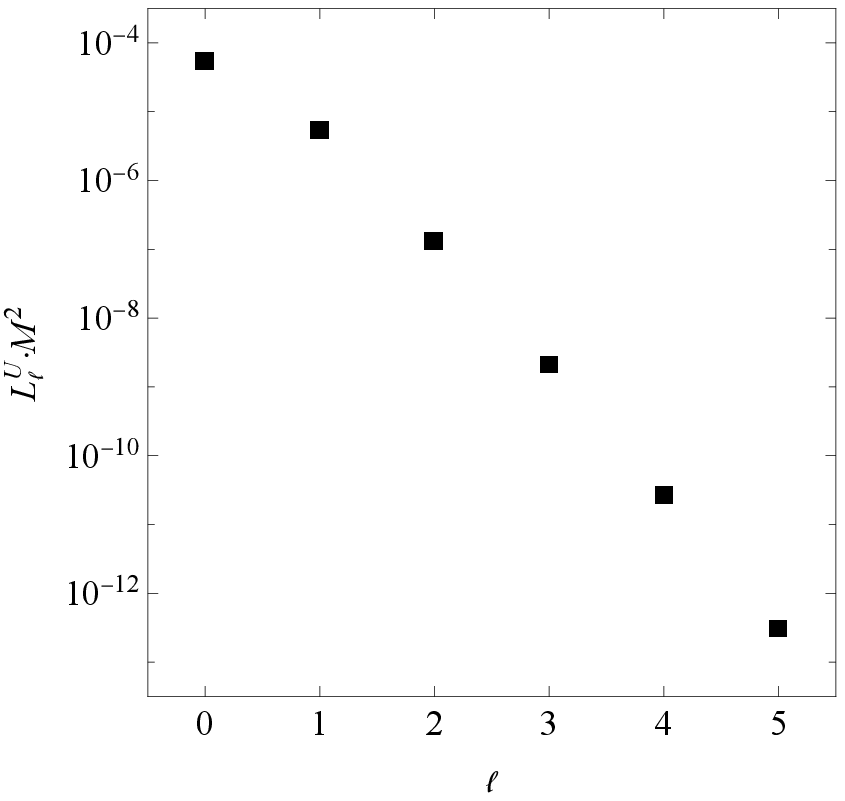}
\end{centering}
\caption{\label{fig:PMR}\emph{(Left panel)} Numerically computed values of the integrand $\hat{E}^{\text{U}}_{\omega\ell}$ from Eq.~\eqref{eq:TuuU} for the ${\ell=0}$ (blue) and ${\ell=1}$ (orange) modes. The area under each of these curves (which approaches the constant $\beta$ as ${\ell\to\infty}$) is used to calculate the Unruh-state renormalized stress-energy component $\langle T_{uu}\rangle^{\text{ren}}_{\text{U}}$ evaluated at the inner horizon.
\\\emph{(Right panel)} Numerically computed values of luminosity $\ell$-modes $L_{\ell}^{\text{U}}$ from Eq.~\eqref{eq:hawkingoutflux}. The sum of all these values from ${\ell=0}$ to ${\ell=\infty}$ yields the Hawking outflux ${4\pi r_-^2\left(\langle T_{uu}\rangle^{\text{ren}}_{\text{U}}-\langle T_{vv}\rangle^{\text{ren}}_{\text{U}}\right)}$ of Eq.~\eqref{eq:TvvU}. All modes are positive and drop to zero exponentially as $\ell$ increases. The fact that the sum over all these points yields a non-zero value indicates that at least one (Kruskalized) component of the renormalized stress-energy tensor diverges at the inner horizon of spherical inner-extremal regular black holes. The constants used for both the left and right panels are ${\alpha=1}$, ${l/M=1/100}$, and ${a_2/M^2=1/10}$.}
\end{figure*}

Once these scattering coefficients are computed numerically for each set of modes specified by $\omega$ and $\ell$, the quantity $\hat{E}^{\text{U}}_{\omega\ell}$ from Eq.~\eqref{eq:TuuU} can be integrated over a sampled set of frequencies with the help of third-order Hermite interpolation built into the software package \texttt{Mathematica}. In practice, instead of sampling points all the way out to ${\omega\to\infty}$, computations of the integrand $\hat{E}^{\text{U}}_{\omega\ell}$ are terminated once it enters deep into the regime in which it decays as ${\omega\cdot\text{e}^{-\omega/k}}$ for some positive $k$, after which the integrand is analytically extended to infinity with the appropriate extrapolation. The values of this integrand for the ${\ell=0}$ and ${\ell=1}$ modes are shown in the left panel of Fig.~\ref{fig:PMR}.

For numerical computations, the outer and inner horizons are chosen to lie at the following radii:
\begin{equation}
    r_+=2M,\qquad r_-=l\left(1+\alpha\frac{l}{M}+\mathcal{O}\left(\frac{l^2}{M^2}\right)\right),
\end{equation}
where $M$ is the mass of the black hole, $\alpha$ is an order-unity parameter, and $l$ is a regularization length scale often identified as the Planck length, where semiclassical gravity breaks down (though it should be noted that for the present choice of constants such an identification cannot be upheld as it would imply that the black hole weighs less than a single grain of sand). The numerical computations done here use the same choices for these constants as in Ref.~\cite{car22}: ${\alpha=1}$ and ${M/l=100}$.

In the left panel of Fig.~\ref{fig:PMR}, though the spectra for only the lowest two angular modes (${\ell=0}$ and ${\ell=1}$) are shown, all higher-$\ell$ modes appear visually similar to the ${\ell=1}$ spectrum on that plot, as the integrated spectrum quickly plateaus to the value $\beta$ as $\ell$ is increased. This constant is numerically found to equal approximately ${1.4\times10^4M^{-4}}$, which is consistent with the parameter range one might expect from an analysis of Reissner-Nordstr\"om black holes (in particular, the inner-extremal $\beta$ for this choice of parameters coincides with the Reissner-Nordstr\"om $\beta$ for a black hole with charge-to-mass ratio ${Q/M\approx0.427}$). As a result, the $uu$-component of the Unruh-state renormalized stress-energy tensor at the inner horizon from Eq.~\eqref{eq:TuuU} is calculated to be ${8.9\times10^5M^{-4}}$. Since this value is non-zero, the corresponding stress-energy component for a set of coordinates that are regular through the inner horizon, such as the Kruskal coordinates of Eq.~\eqref{eq:UV}, will diverge.

Since the inner-extremal regular black hole spacetime under consideration here is spherically symmetric and static, energy-momentum conservation implies that in spherically symmetric, static quantum states like the Unruh state, the quantity
\begin{equation}
    4\pi r^2\Big(\langle T_{uu}\rangle^{\text{ren}}_{\text{U}}-\langle T_{vv}\rangle^{\text{ren}}_{\text{U}}\Big)
\end{equation}
must be conserved everywhere in the spacetime \cite{zil20}. For some vacuum states like the Hartle-Hawking state, this constant trivially vanishes, but for the Unruh state, it can be evaluated at the inner horizon as the sum ${\sum_{\ell=0}^{\infty}L_{\ell}^{\text{U}}}$ from Eqs.~\eqref{eq:TvvU} and \eqref{eq:hawkingoutflux}. For the Unruh state, this luminosity coincides precisely with the Hawking outflux. For the choice of inner-extremal parameters used throughout this section, the computed Hawking outflux equals ${6.81142\times10^{-5}M^{-2}}$. To obtain this value, similar to the methodologies described above, the external scattering coefficient is sampled for a set of frequencies and extrapolated with the knowledge that at high frequencies, the integrand of Eq.~\eqref{eq:hawkingoutflux} behaves as ${\omega\cdot\text{e}^{-2\widetilde{\omega}}}$, while at low frequencies, it behaves as a power law in $\omega$. Then, the spectrum is integrated over all frequencies and summed over larger and larger values of $\ell$ until convergence is reached, as shown in the right panel of Fig.~\ref{fig:PMR}.

The fact that the Hawking outflux does not vanish at the inner horizon indicates that ${\langle T_{uu}\rangle^{\text{ren}}_{\text{U}}}$ and ${\langle T_{vv}\rangle^{\text{ren}}_{\text{U}}}$ can never simultaneously equal zero and therefore that at least one component (in coordinates that are regular across the inner horizon) of the renormalized stress-energy tensor will always diverge there. The remarkable aspect of this result is that the semiclassical divergence occurs regardless of anything happening in the interior, such as a vanishing surface gravity at the inner horizon or some anomalous scattering governed by $A_{\omega\ell}$ and $B_{\omega\ell}$. Rather, from Eq.~\eqref{eq:hawkingoutflux}, this divergence depends only on the external portion of the spacetime, characterized by the outer horizon's surface gravity ${\kappa(r_+)}$ and the external transmission coefficient ${\tau_{\omega\ell}^{\text{up}}}$.

\section{Outlook}
\label{sec:out}
In the absence of a full theory of quantum gravity, one may hope that using an effective field theory to describe the semiclassical behavior of gravity (valid up the the Planck energy) would be enough to provide a complete model of astrophysical black holes formed from collapse. If this were true, one should be able to write down a completely classical, singularity-free metric to describe the black hole, with some contributions from both classical and semiclassical sources via Eq.~\eqref{eq:semi_einstein}. The inner-extremal regular black hole metrics of Eqs.~\eqref{eq:SSS} and \eqref{eq:lineelement_RIERBH} are two potential classes of such models, especially promising due to their avoidance of the classical mass inflation instability.

The key takeaway of the present analysis is that for black holes formed from astrophysical collapse, no regular black hole models with an inner horizon will be semiclassically stable and regular, regardless of whether or not the inner horizon is fine-tuned so that its surface gravity vanishes (like in the inner-extremal models). An Unruh-state semiclassical divergence at the inner horizon is driven by both the inner and outer horizons' surface gravities, so that the only singularity-free black holes models that can avoid the semiclassical instability are extremal black holes.

The semiclassical divergence present at the inner horizon of inner-extremal regular black holes has here been demonstrated with the calculation of several different important semiclassical quantities. First, the effective Hawking temperature $\kappa_{\text{eff}}$ was calculated for inertial ingoing and outgoing observers passing through the inner horizon, for both spherical (Sec.~\ref{subsec:sph}) and rotating (Sec.~\ref{subsec:rot}) inner-extremal regular black holes. The effective temperature was found to diverge as ${(r-r_-)^{-1}}$ for outgoing observers at the inner horizon looking up at the sky above and as ${(r-r_-)^{-3}}$ for ingoing observers at the inner horizon looking down at the horizon below (the factor of 3 corresponds to the number of degenerate inner horizons, equal to the lowest non-zero order in a local expansion of the horizon function $\Delta$).

Second, the renormalized stress-energy tensor for a massless, scalar field in the spherical inner-extremal geometry has been calculated in Sec.~\ref{subsec:pol} using the Polyakov approximation (i.e.\ averaging over the angular degrees of freedom so that an exact answer can be obtained in 1+1 dimensions). The normal stress component of this tensor in outgoing Eddington-Finkelstein coordinates (${\langle T_{uu}\rangle^{\text{ren}}}$) remains non-zero at the inner horizon in both the Unruh and Minkowski ``in'' vacuum states, which indicates that the physical stress-energy will diverge when one transforms to a set of coordinates that are regular across that surface.

Finally, to confirm that the 1+1D calculations of Sec.~\ref{subsec:pol} are not missing any crucial information from the scattering of higher-$\ell$ angular modes in the full 3+1 dimensions, the renormalized stress-energy tensor has been calculated numerically for a specific choice of parameters in Sec.~\ref{subsec:pmr} using pragmatic mode-sum renormalization. To do so requires finding the exterior and interior scattering coefficients for free waves traveling from infinity to the outer horizon and from the outer horizon to the inner horizon, respectively. The result is the same as in the 1+1D case: the renormalized stress-energy in outgoing Eddington-Finkelstein coordinates do not vanish at the inner horizon, so that a semiclassical singularity will emerge there if the spacetime remains static. This divergence will always occur for at least one leg of the inner horizon, since the difference ${\langle T_{uu}\rangle^{\text{ren}}_{\text{U}}-\langle T_{vv}\rangle^{\text{ren}}_{\text{U}}}$ in the Unruh state is always proportional to the non-zero Hawking outflux.

It would thus appear that any semiclassically self-consistent model of a regular black hole one may come up with cannot have an inner horizon that is spatially separated from the outer horizon, no matter how degenerate it may be. It would be interesting to analyze how the semiclassical back-reaction dynamically affects the inner-extremal geometry if the constraints of staticity are relaxed\textemdash the inner horizon may evaporate outward to meet the outer horizon and perhaps evolve to a new, non-black-hole geometry, for example. However, the vanishing of ${\langle T_{vv}\rangle^{\text{ren}}}$ in Eqs.~\eqref{eq:Tmunuin_rm} and \eqref{eq:TmunuU} at the inner horizon offers an indication that forcing the inner horizon's surface gravity to vanish only strengthens the semiclassical divergence, since it is precisely this surface gravity that would cause the inner horizon to evaporate. Instead, it is likely that the semiclassical inflation near the inner horizon will occur too rapidly for the geometry to have time to react, so that a curvature singularity forms. One must then appeal to higher-order theories of quantum gravity to understand how spacetime evolves further \cite{cro22}.

\bibliography{apsbib}

\providecommand{\noopsort}[1]{}\providecommand{\singleletter}[1]{#1}%
\begin{thebibliography}{45}%
\makeatletter
\providecommand \@ifxundefined [1]{%
 \@ifx{#1\undefined}
}%
\providecommand \@ifnum [1]{%
 \ifnum #1\expandafter \@firstoftwo
 \else \expandafter \@secondoftwo
 \fi
}%
\providecommand \@ifx [1]{%
 \ifx #1\expandafter \@firstoftwo
 \else \expandafter \@secondoftwo
 \fi
}%
\providecommand \natexlab [1]{#1}%
\providecommand \enquote  [1]{``#1''}%
\providecommand \bibnamefont  [1]{#1}%
\providecommand \bibfnamefont [1]{#1}%
\providecommand \citenamefont [1]{#1}%
\providecommand \href@noop [0]{\@secondoftwo}%
\providecommand \href [0]{\begingroup \@sanitize@url \@href}%
\providecommand \@href[1]{\@@startlink{#1}\@@href}%
\providecommand \@@href[1]{\endgroup#1\@@endlink}%
\providecommand \@sanitize@url [0]{\catcode `\\12\catcode `\$12\catcode
  `\&12\catcode `\#12\catcode `\^12\catcode `\_12\catcode `\%12\relax}%
\providecommand \@@startlink[1]{}%
\providecommand \@@endlink[0]{}%
\providecommand \url  [0]{\begingroup\@sanitize@url \@url }%
\providecommand \@url [1]{\endgroup\@href {#1}{\urlprefix }}%
\providecommand \urlprefix  [0]{URL }%
\providecommand \Eprint [0]{\href }%
\providecommand \doibase [0]{https://doi.org/}%
\providecommand \selectlanguage [0]{\@gobble}%
\providecommand \bibinfo  [0]{\@secondoftwo}%
\providecommand \bibfield  [0]{\@secondoftwo}%
\providecommand \translation [1]{[#1]}%
\providecommand \BibitemOpen [0]{}%
\providecommand \bibitemStop [0]{}%
\providecommand \bibitemNoStop [0]{.\EOS\space}%
\providecommand \EOS [0]{\spacefactor3000\relax}%
\providecommand \BibitemShut  [1]{\csname bibitem#1\endcsname}%
\let\auto@bib@innerbib\@empty
\bibitem [{\citenamefont {Penrose}(1965)}]{pen65}%
  \BibitemOpen
  \bibfield  {author} {\bibinfo {author} {\bibfnamefont {R.}~\bibnamefont
  {Penrose}},\ }\href {https://doi.org/10.1103/PhysRevLett.14.57} {\bibfield
  {journal} {\bibinfo  {journal} {Phys. Rev. Lett.}\ }\textbf {\bibinfo
  {volume} {14}},\ \bibinfo {pages} {57} (\bibinfo {year} {1965})}\BibitemShut
  {NoStop}%
\bibitem [{\citenamefont {Dymnikova}(2002)}]{dym02}%
  \BibitemOpen
  \bibfield  {author} {\bibinfo {author} {\bibfnamefont {I.}~\bibnamefont
  {Dymnikova}},\ }\href {https://doi.org/10.1088/0264-9381/19/4/306} {\bibfield
   {journal} {\bibinfo  {journal} {Classical and Quantum Gravity}\ }\textbf
  {\bibinfo {volume} {19}},\ \bibinfo {pages} {725} (\bibinfo {year}
  {2002})}\BibitemShut {NoStop}%
\bibitem [{\citenamefont {Carballo-Rubio}\ \emph
  {et~al.}(2020{\natexlab{a}})\citenamefont {Carballo-Rubio}, \citenamefont
  {Di~Filippo}, \citenamefont {Liberati},\ and\ \citenamefont
  {Visser}}]{car20a}%
  \BibitemOpen
  \bibfield  {author} {\bibinfo {author} {\bibfnamefont {R.}~\bibnamefont
  {Carballo-Rubio}}, \bibinfo {author} {\bibfnamefont {F.}~\bibnamefont
  {Di~Filippo}}, \bibinfo {author} {\bibfnamefont {S.}~\bibnamefont
  {Liberati}},\ and\ \bibinfo {author} {\bibfnamefont {M.}~\bibnamefont
  {Visser}},\ }\href {https://doi.org/10.1103/PhysRevD.101.084047} {\bibfield
  {journal} {\bibinfo  {journal} {Phys. Rev. D}\ }\textbf {\bibinfo {volume}
  {101}},\ \bibinfo {pages} {084047} (\bibinfo {year}
  {2020}{\natexlab{a}})}\BibitemShut {NoStop}%
\bibitem [{\citenamefont {Carballo-Rubio}\ \emph
  {et~al.}(2020{\natexlab{b}})\citenamefont {Carballo-Rubio}, \citenamefont
  {Filippo}, \citenamefont {Liberati},\ and\ \citenamefont {Visser}}]{car20b}%
  \BibitemOpen
  \bibfield  {author} {\bibinfo {author} {\bibfnamefont {R.}~\bibnamefont
  {Carballo-Rubio}}, \bibinfo {author} {\bibfnamefont {F.~D.}\ \bibnamefont
  {Filippo}}, \bibinfo {author} {\bibfnamefont {S.}~\bibnamefont {Liberati}},\
  and\ \bibinfo {author} {\bibfnamefont {M.}~\bibnamefont {Visser}},\ }\href
  {https://doi.org/10.1088/1361-6382/ab8141} {\bibfield  {journal} {\bibinfo
  {journal} {Classical and Quantum Gravity}\ }\textbf {\bibinfo {volume}
  {37}},\ \bibinfo {pages} {145005} (\bibinfo {year}
  {2020}{\natexlab{b}})}\BibitemShut {NoStop}%
\bibitem [{\citenamefont {Penrose}(1968)}]{pen68}%
  \BibitemOpen
  \bibfield  {author} {\bibinfo {author} {\bibfnamefont {R.}~\bibnamefont
  {Penrose}},\ }in\ \href@noop {} {\emph {\bibinfo {booktitle} {{Battelle
  {R}encontres: 1967 lectures in mathematics and physics}}}},\ \bibinfo
  {editor} {edited by\ \bibinfo {editor} {\bibfnamefont {C.}~\bibnamefont
  {de~Witt-Morette}}\ and\ \bibinfo {editor} {\bibfnamefont {J.~A.}\
  \bibnamefont {Wheeler}}}\ (\bibinfo  {publisher} {W. A. Benjamin},\ \bibinfo
  {address} {New York},\ \bibinfo {year} {1968})\ pp.\ \bibinfo {pages}
  {121--235}\BibitemShut {NoStop}%
\bibitem [{\citenamefont {Simpson}\ and\ \citenamefont
  {Penrose}(1973)}]{sim73}%
  \BibitemOpen
  \bibfield  {author} {\bibinfo {author} {\bibfnamefont {M.}~\bibnamefont
  {Simpson}}\ and\ \bibinfo {author} {\bibfnamefont {R.}~\bibnamefont
  {Penrose}},\ }\href {https://doi.org/10.1007/BF00792069} {\bibfield
  {journal} {\bibinfo  {journal} {Int.\ J.\ Theor.\ Phys.}\ }\textbf {\bibinfo
  {volume} {7}},\ \bibinfo {pages} {183} (\bibinfo {year} {1973})}\BibitemShut
  {NoStop}%
\bibitem [{\citenamefont {Poisson}\ and\ \citenamefont {Israel}(1990)}]{poi90}%
  \BibitemOpen
  \bibfield  {author} {\bibinfo {author} {\bibfnamefont {E.}~\bibnamefont
  {Poisson}}\ and\ \bibinfo {author} {\bibfnamefont {W.}~\bibnamefont
  {Israel}},\ }\href {https://doi.org/10.1103/PhysRevD.41.1796} {\bibfield
  {journal} {\bibinfo  {journal} {Phys. Rev. D}\ }\textbf {\bibinfo {volume}
  {41}},\ \bibinfo {pages} {1796} (\bibinfo {year} {1990})}\BibitemShut
  {NoStop}%
\bibitem [{\citenamefont {Ori}(1991)}]{ori91}%
  \BibitemOpen
  \bibfield  {author} {\bibinfo {author} {\bibfnamefont {A.}~\bibnamefont
  {Ori}},\ }\href {https://doi.org/10.1103/PhysRevLett.67.789} {\bibfield
  {journal} {\bibinfo  {journal} {Phys. Rev. Lett.}\ }\textbf {\bibinfo
  {volume} {67}},\ \bibinfo {pages} {789} (\bibinfo {year} {1991})}\BibitemShut
  {NoStop}%
\bibitem [{\citenamefont {Hamilton}\ and\ \citenamefont
  {Avelino}(2010)}]{ham10}%
  \BibitemOpen
  \bibfield  {author} {\bibinfo {author} {\bibfnamefont {A.~J.}\ \bibnamefont
  {Hamilton}}\ and\ \bibinfo {author} {\bibfnamefont {P.~P.}\ \bibnamefont
  {Avelino}},\ }\href {https://doi.org//10.1016/j.physrep.2010.06.002}
  {\bibfield  {journal} {\bibinfo  {journal} {Physics Reports}\ }\textbf
  {\bibinfo {volume} {495}},\ \bibinfo {pages} {1 } (\bibinfo {year}
  {2010})}\BibitemShut {NoStop}%
\bibitem [{\citenamefont {Carballo-Rubio}\ \emph {et~al.}(2022)\citenamefont
  {Carballo-Rubio}, \citenamefont {Di~Filippo}, \citenamefont {Liberati},
  \citenamefont {Pacilio},\ and\ \citenamefont {Visser}}]{car22}%
  \BibitemOpen
  \bibfield  {author} {\bibinfo {author} {\bibfnamefont {R.}~\bibnamefont
  {Carballo-Rubio}}, \bibinfo {author} {\bibfnamefont {F.}~\bibnamefont
  {Di~Filippo}}, \bibinfo {author} {\bibfnamefont {S.}~\bibnamefont
  {Liberati}}, \bibinfo {author} {\bibfnamefont {C.}~\bibnamefont {Pacilio}},\
  and\ \bibinfo {author} {\bibfnamefont {M.}~\bibnamefont {Visser}},\ }\href
  {https://doi.org/10.1007/JHEP09(2022)118} {\bibfield  {journal} {\bibinfo
  {journal} {Journal of High Energy Physics}\ }\textbf {\bibinfo {volume}
  {2022}},\ \bibinfo {pages} {118} (\bibinfo {year} {2022})}\BibitemShut
  {NoStop}%
\bibitem [{\citenamefont {Franzin}\ \emph {et~al.}(2022)\citenamefont
  {Franzin}, \citenamefont {Liberati}, \citenamefont {Mazza},\ and\
  \citenamefont {Vellucci}}]{fra22}%
  \BibitemOpen
  \bibfield  {author} {\bibinfo {author} {\bibfnamefont {E.}~\bibnamefont
  {Franzin}}, \bibinfo {author} {\bibfnamefont {S.}~\bibnamefont {Liberati}},
  \bibinfo {author} {\bibfnamefont {J.}~\bibnamefont {Mazza}},\ and\ \bibinfo
  {author} {\bibfnamefont {V.}~\bibnamefont {Vellucci}},\ }\href
  {https://doi.org/10.1103/PhysRevD.106.104060} {\bibfield  {journal} {\bibinfo
   {journal} {Phys. Rev. D}\ }\textbf {\bibinfo {volume} {106}},\ \bibinfo
  {pages} {104060} (\bibinfo {year} {2022})}\BibitemShut {NoStop}%
\bibitem [{\citenamefont {Maeda}\ \emph {et~al.}(2005)\citenamefont {Maeda},
  \citenamefont {Torii},\ and\ \citenamefont {Harada}}]{mae05}%
  \BibitemOpen
  \bibfield  {author} {\bibinfo {author} {\bibfnamefont {H.}~\bibnamefont
  {Maeda}}, \bibinfo {author} {\bibfnamefont {T.}~\bibnamefont {Torii}},\ and\
  \bibinfo {author} {\bibfnamefont {T.}~\bibnamefont {Harada}},\ }\href
  {https://doi.org/10.1103/PhysRevD.71.064015} {\bibfield  {journal} {\bibinfo
  {journal} {Phys. Rev. D}\ }\textbf {\bibinfo {volume} {71}},\ \bibinfo
  {pages} {064015} (\bibinfo {year} {2005})}\BibitemShut {NoStop}%
\bibitem [{\citenamefont {Frolov}\ and\ \citenamefont
  {Zelnikov}(2017{\natexlab{a}})}]{fro17b}%
  \BibitemOpen
  \bibfield  {author} {\bibinfo {author} {\bibfnamefont {V.~P.}\ \bibnamefont
  {Frolov}}\ and\ \bibinfo {author} {\bibfnamefont {A.}~\bibnamefont
  {Zelnikov}},\ }\href {https://doi.org/10.1103/PhysRevD.95.124028} {\bibfield
  {journal} {\bibinfo  {journal} {Phys. Rev. D}\ }\textbf {\bibinfo {volume}
  {95}},\ \bibinfo {pages} {124028} (\bibinfo {year}
  {2017}{\natexlab{a}})}\BibitemShut {NoStop}%
\bibitem [{\citenamefont {Bonanno}\ \emph {et~al.}(2023)\citenamefont
  {Bonanno}, \citenamefont {Khosravi},\ and\ \citenamefont
  {Saueressig}}]{bon23}%
  \BibitemOpen
  \bibfield  {author} {\bibinfo {author} {\bibfnamefont {A.}~\bibnamefont
  {Bonanno}}, \bibinfo {author} {\bibfnamefont {A.-P.}\ \bibnamefont
  {Khosravi}},\ and\ \bibinfo {author} {\bibfnamefont {F.}~\bibnamefont
  {Saueressig}},\ }\href {https://doi.org/10.1103/PhysRevD.107.024005}
  {\bibfield  {journal} {\bibinfo  {journal} {Phys. Rev. D}\ }\textbf {\bibinfo
  {volume} {107}},\ \bibinfo {pages} {024005} (\bibinfo {year}
  {2023})}\BibitemShut {NoStop}%
\bibitem [{\citenamefont {Frolov}\ and\ \citenamefont
  {Zelnikov}(2017{\natexlab{b}})}]{fro17a}%
  \BibitemOpen
  \bibfield  {author} {\bibinfo {author} {\bibfnamefont {V.~P.}\ \bibnamefont
  {Frolov}}\ and\ \bibinfo {author} {\bibfnamefont {A.}~\bibnamefont
  {Zelnikov}},\ }\href {https://doi.org/10.1103/PhysRevD.95.044042} {\bibfield
  {journal} {\bibinfo  {journal} {Phys. Rev. D}\ }\textbf {\bibinfo {volume}
  {95}},\ \bibinfo {pages} {044042} (\bibinfo {year}
  {2017}{\natexlab{b}})}\BibitemShut {NoStop}%
\bibitem [{\citenamefont {Zilberman}\ \emph {et~al.}(2020)\citenamefont
  {Zilberman}, \citenamefont {Levi},\ and\ \citenamefont {Ori}}]{zil20}%
  \BibitemOpen
  \bibfield  {author} {\bibinfo {author} {\bibfnamefont {N.}~\bibnamefont
  {Zilberman}}, \bibinfo {author} {\bibfnamefont {A.}~\bibnamefont {Levi}},\
  and\ \bibinfo {author} {\bibfnamefont {A.}~\bibnamefont {Ori}},\ }\href
  {https://doi.org/10.1103/PhysRevLett.124.171302} {\bibfield  {journal}
  {\bibinfo  {journal} {Phys. Rev. Lett.}\ }\textbf {\bibinfo {volume} {124}},\
  \bibinfo {pages} {171302} (\bibinfo {year} {2020})}\BibitemShut {NoStop}%
\bibitem [{\citenamefont {Hollands}\ \emph {et~al.}(2020)\citenamefont
  {Hollands}, \citenamefont {Wald},\ and\ \citenamefont {Zahn}}]{hol20}%
  \BibitemOpen
  \bibfield  {author} {\bibinfo {author} {\bibfnamefont {S.}~\bibnamefont
  {Hollands}}, \bibinfo {author} {\bibfnamefont {R.~M.}\ \bibnamefont {Wald}},\
  and\ \bibinfo {author} {\bibfnamefont {J.}~\bibnamefont {Zahn}},\ }\href
  {https://doi.org/10.1088/1361-6382/ab8052} {\bibfield  {journal} {\bibinfo
  {journal} {Class. Quant. Grav.}\ }\textbf {\bibinfo {volume} {37}},\ \bibinfo
  {pages} {115009} (\bibinfo {year} {2020})}\BibitemShut {NoStop}%
\bibitem [{\citenamefont {Hiscock}(1980)}]{his80}%
  \BibitemOpen
  \bibfield  {author} {\bibinfo {author} {\bibfnamefont {W.~A.}\ \bibnamefont
  {Hiscock}},\ }\href {https://doi.org/10.1103/PhysRevD.21.2057} {\bibfield
  {journal} {\bibinfo  {journal} {Phys. Rev. D}\ }\textbf {\bibinfo {volume}
  {21}},\ \bibinfo {pages} {2057} (\bibinfo {year} {1980})}\BibitemShut
  {NoStop}%
\bibitem [{\citenamefont {Zilberman}\ \emph {et~al.}(2022)\citenamefont
  {Zilberman}, \citenamefont {Casals}, \citenamefont {Ori},\ and\ \citenamefont
  {Ottewill}}]{zil22b}%
  \BibitemOpen
  \bibfield  {author} {\bibinfo {author} {\bibfnamefont {N.}~\bibnamefont
  {Zilberman}}, \bibinfo {author} {\bibfnamefont {M.}~\bibnamefont {Casals}},
  \bibinfo {author} {\bibfnamefont {A.}~\bibnamefont {Ori}},\ and\ \bibinfo
  {author} {\bibfnamefont {A.~C.}\ \bibnamefont {Ottewill}},\ }\href
  {https://doi.org/10.1103/PhysRevLett.129.261102} {\bibfield  {journal}
  {\bibinfo  {journal} {Phys. Rev. Lett.}\ }\textbf {\bibinfo {volume} {129}},\
  \bibinfo {pages} {261102} (\bibinfo {year} {2022})}\BibitemShut {NoStop}%
\bibitem [{\citenamefont {Barcel\'o}\ \emph {et~al.}(2022)\citenamefont
  {Barcel\'o}, \citenamefont {Boyanov}, \citenamefont {Carballo-Rubio},\ and\
  \citenamefont {Garay}}]{bar22}%
  \BibitemOpen
  \bibfield  {author} {\bibinfo {author} {\bibfnamefont {C.}~\bibnamefont
  {Barcel\'o}}, \bibinfo {author} {\bibfnamefont {V.}~\bibnamefont {Boyanov}},
  \bibinfo {author} {\bibfnamefont {R.}~\bibnamefont {Carballo-Rubio}},\ and\
  \bibinfo {author} {\bibfnamefont {L.~J.}\ \bibnamefont {Garay}},\ }\href
  {https://doi.org/10.1103/PhysRevD.106.124006} {\bibfield  {journal} {\bibinfo
   {journal} {Phys. Rev. D}\ }\textbf {\bibinfo {volume} {106}},\ \bibinfo
  {pages} {124006} (\bibinfo {year} {2022})}\BibitemShut {NoStop}%
\bibitem [{\citenamefont {Bonanno}\ \emph {et~al.}(2021)\citenamefont
  {Bonanno}, \citenamefont {Khosravi},\ and\ \citenamefont
  {Saueressig}}]{bon21}%
  \BibitemOpen
  \bibfield  {author} {\bibinfo {author} {\bibfnamefont {A.}~\bibnamefont
  {Bonanno}}, \bibinfo {author} {\bibfnamefont {A.-P.}\ \bibnamefont
  {Khosravi}},\ and\ \bibinfo {author} {\bibfnamefont {F.}~\bibnamefont
  {Saueressig}},\ }\href {https://doi.org/10.1103/PhysRevD.103.124027}
  {\bibfield  {journal} {\bibinfo  {journal} {Phys. Rev. D}\ }\textbf {\bibinfo
  {volume} {103}},\ \bibinfo {pages} {124027} (\bibinfo {year}
  {2021})}\BibitemShut {NoStop}%
\bibitem [{\citenamefont {Barcel{\'{o}}}\ \emph {et~al.}(2021)\citenamefont
  {Barcel{\'{o}}}, \citenamefont {Boyanov}, \citenamefont {Carballo-Rubio},\
  and\ \citenamefont {Garay}}]{bar21}%
  \BibitemOpen
  \bibfield  {author} {\bibinfo {author} {\bibfnamefont {C.}~\bibnamefont
  {Barcel{\'{o}}}}, \bibinfo {author} {\bibfnamefont {V.}~\bibnamefont
  {Boyanov}}, \bibinfo {author} {\bibfnamefont {R.}~\bibnamefont
  {Carballo-Rubio}},\ and\ \bibinfo {author} {\bibfnamefont {L.~J.}\
  \bibnamefont {Garay}},\ }\href {https://doi.org/10.1088/1361-6382/abf89c}
  {\bibfield  {journal} {\bibinfo  {journal} {Class. Quant. Grav.}\ }\textbf
  {\bibinfo {volume} {38}},\ \bibinfo {pages} {125003} (\bibinfo {year}
  {2021})}\BibitemShut {NoStop}%
\bibitem [{\citenamefont {McMaken}\ and\ \citenamefont
  {Hamilton}(2023)}]{mcm23}%
  \BibitemOpen
  \bibfield  {author} {\bibinfo {author} {\bibfnamefont {T.}~\bibnamefont
  {McMaken}}\ and\ \bibinfo {author} {\bibfnamefont {A.~J.~S.}\ \bibnamefont
  {Hamilton}},\ }\href {https://doi.org/10.1103/PhysRevD.107.085010} {\bibfield
   {journal} {\bibinfo  {journal} {Phys. Rev. D}\ }\textbf {\bibinfo {volume}
  {107}},\ \bibinfo {pages} {085010} (\bibinfo {year} {2023})}\BibitemShut
  {NoStop}%
\bibitem [{\citenamefont {Barbado}\ \emph {et~al.}(2016)\citenamefont
  {Barbado}, \citenamefont {Barcel\'o}, \citenamefont {Garay},\ and\
  \citenamefont {Jannes}}]{bar16}%
  \BibitemOpen
  \bibfield  {author} {\bibinfo {author} {\bibfnamefont {L.~C.}\ \bibnamefont
  {Barbado}}, \bibinfo {author} {\bibfnamefont {C.}~\bibnamefont {Barcel\'o}},
  \bibinfo {author} {\bibfnamefont {L.~J.}\ \bibnamefont {Garay}},\ and\
  \bibinfo {author} {\bibfnamefont {G.}~\bibnamefont {Jannes}},\ }\href
  {https://doi.org/10.1103/PhysRevD.94.064004} {\bibfield  {journal} {\bibinfo
  {journal} {Phys. Rev. D}\ }\textbf {\bibinfo {volume} {94}},\ \bibinfo
  {pages} {064004} (\bibinfo {year} {2016})}\BibitemShut {NoStop}%
\bibitem [{\citenamefont {Ju\'{a}rez-Aubry}\ and\ \citenamefont
  {Louko}(2022)}]{jua22}%
  \BibitemOpen
  \bibfield  {author} {\bibinfo {author} {\bibfnamefont {B.~A.}\ \bibnamefont
  {Ju\'{a}rez-Aubry}}\ and\ \bibinfo {author} {\bibfnamefont {J.}~\bibnamefont
  {Louko}},\ }\href {https://doi.org/10.1116/5.0073373} {\bibfield  {journal}
  {\bibinfo  {journal} {AVS Quantum Science}\ }\textbf {\bibinfo {volume}
  {4}},\ \bibinfo {pages} {013201} (\bibinfo {year} {2022})}\BibitemShut
  {NoStop}%
\bibitem [{\citenamefont {Hawking}(1975)}]{haw75}%
  \BibitemOpen
  \bibfield  {author} {\bibinfo {author} {\bibfnamefont {S.~W.}\ \bibnamefont
  {Hawking}},\ }\href {https://doi.org/10.1007/BF02345020} {\bibfield
  {journal} {\bibinfo  {journal} {Communications in Mathematical Physics}\
  }\textbf {\bibinfo {volume} {43}},\ \bibinfo {pages} {199} (\bibinfo {year}
  {1975})}\BibitemShut {NoStop}%
\bibitem [{\citenamefont {Barcel{\'o}}\ \emph {et~al.}(2011)\citenamefont
  {Barcel{\'o}}, \citenamefont {Liberati}, \citenamefont {Sonego},\ and\
  \citenamefont {Visser}}]{bar11a}%
  \BibitemOpen
  \bibfield  {author} {\bibinfo {author} {\bibfnamefont {C.}~\bibnamefont
  {Barcel{\'o}}}, \bibinfo {author} {\bibfnamefont {S.}~\bibnamefont
  {Liberati}}, \bibinfo {author} {\bibfnamefont {S.}~\bibnamefont {Sonego}},\
  and\ \bibinfo {author} {\bibfnamefont {M.}~\bibnamefont {Visser}},\ }\href
  {https://doi.org/10.1007/JHEP02(2011)003} {\bibfield  {journal} {\bibinfo
  {journal} {Journal of High Energy Physics}\ }\textbf {\bibinfo {volume}
  {2011}},\ \bibinfo {pages} {1} (\bibinfo {year} {2011})}\BibitemShut
  {NoStop}%
\bibitem [{\citenamefont {Barcel\'o}\ \emph {et~al.}(2011)\citenamefont
  {Barcel\'o}, \citenamefont {Liberati}, \citenamefont {Sonego},\ and\
  \citenamefont {Visser}}]{bar11b}%
  \BibitemOpen
  \bibfield  {author} {\bibinfo {author} {\bibfnamefont {C.}~\bibnamefont
  {Barcel\'o}}, \bibinfo {author} {\bibfnamefont {S.}~\bibnamefont {Liberati}},
  \bibinfo {author} {\bibfnamefont {S.}~\bibnamefont {Sonego}},\ and\ \bibinfo
  {author} {\bibfnamefont {M.}~\bibnamefont {Visser}},\ }\href
  {https://doi.org/10.1103/PhysRevD.83.041501} {\bibfield  {journal} {\bibinfo
  {journal} {\prd}\ }\textbf {\bibinfo {volume} {83}},\ \bibinfo {pages}
  {041501} (\bibinfo {year} {2011})}\BibitemShut {NoStop}%
\bibitem [{\citenamefont {Unruh}(1976)}]{unr76}%
  \BibitemOpen
  \bibfield  {author} {\bibinfo {author} {\bibfnamefont {W.~G.}\ \bibnamefont
  {Unruh}},\ }\href {https://doi.org/10.1103/PhysRevD.14.870} {\bibfield
  {journal} {\bibinfo  {journal} {Phys. Rev. D}\ }\textbf {\bibinfo {volume}
  {14}},\ \bibinfo {pages} {870} (\bibinfo {year} {1976})}\BibitemShut
  {NoStop}%
\bibitem [{\citenamefont {Hamilton}(2018)}]{ham18}%
  \BibitemOpen
  \bibfield  {author} {\bibinfo {author} {\bibfnamefont {A.~J.~S.}\
  \bibnamefont {Hamilton}},\ }\href {https://doi.org/10.1007/s10714-018-2369-1}
  {\bibfield  {journal} {\bibinfo  {journal} {General Relativity and
  Gravitation}\ }\textbf {\bibinfo {volume} {50}},\ \bibinfo {pages} {50}
  (\bibinfo {year} {2018})}\BibitemShut {NoStop}%
\bibitem [{\citenamefont {Crispino}\ \emph {et~al.}(2016)\citenamefont
  {Crispino}, \citenamefont {Higuchi}, \citenamefont {Oliveira},\ and\
  \citenamefont {de~Oliveira}}]{cri16}%
  \BibitemOpen
  \bibfield  {author} {\bibinfo {author} {\bibfnamefont {L.~C.~B.}\
  \bibnamefont {Crispino}}, \bibinfo {author} {\bibfnamefont {A.}~\bibnamefont
  {Higuchi}}, \bibinfo {author} {\bibfnamefont {L.~A.}\ \bibnamefont
  {Oliveira}},\ and\ \bibinfo {author} {\bibfnamefont {E.~S.}\ \bibnamefont
  {de~Oliveira}},\ }\href {https://doi.org/10.1140/epjc/s10052-016-3972-5}
  {\bibfield  {journal} {\bibinfo  {journal} {The European Physical Journal C}\
  }\textbf {\bibinfo {volume} {76}},\ \bibinfo {pages} {168} (\bibinfo {year}
  {2016})}\BibitemShut {NoStop}%
\bibitem [{\citenamefont {Boyer}\ and\ \citenamefont
  {Lindquist}(1967)}]{boy67}%
  \BibitemOpen
  \bibfield  {author} {\bibinfo {author} {\bibfnamefont {R.~H.}\ \bibnamefont
  {Boyer}}\ and\ \bibinfo {author} {\bibfnamefont {R.~W.}\ \bibnamefont
  {Lindquist}},\ }\href {https://doi.org/10.1063/1.1705193} {\bibfield
  {journal} {\bibinfo  {journal} {Journal of Mathematical Physics}\ }\textbf
  {\bibinfo {volume} {8}},\ \bibinfo {pages} {265} (\bibinfo {year}
  {1967})}\BibitemShut {NoStop}%
\bibitem [{\citenamefont {Carter}(1968)}]{car68}%
  \BibitemOpen
  \bibfield  {author} {\bibinfo {author} {\bibfnamefont {B.}~\bibnamefont
  {Carter}},\ }\href {https://doi.org/10.1103/PhysRev.174.1559} {\bibfield
  {journal} {\bibinfo  {journal} {Phys. Rev.}\ }\textbf {\bibinfo {volume}
  {174}},\ \bibinfo {pages} {1559} (\bibinfo {year} {1968})}\BibitemShut
  {NoStop}%
\bibitem [{\citenamefont {Duff}(1994)}]{duf94}%
  \BibitemOpen
  \bibfield  {author} {\bibinfo {author} {\bibfnamefont {M.~J.}\ \bibnamefont
  {Duff}},\ }\href {https://doi.org/10.1088/0264-9381/11/6/004} {\bibfield
  {journal} {\bibinfo  {journal} {Classical and Quantum Gravity}\ }\textbf
  {\bibinfo {volume} {11}},\ \bibinfo {pages} {1387} (\bibinfo {year}
  {1994})}\BibitemShut {NoStop}%
\bibitem [{\citenamefont {Anderson}\ \emph {et~al.}(2007)\citenamefont
  {Anderson}, \citenamefont {Mottola},\ and\ \citenamefont {Vaulin}}]{and07}%
  \BibitemOpen
  \bibfield  {author} {\bibinfo {author} {\bibfnamefont {P.~R.}\ \bibnamefont
  {Anderson}}, \bibinfo {author} {\bibfnamefont {E.}~\bibnamefont {Mottola}},\
  and\ \bibinfo {author} {\bibfnamefont {R.}~\bibnamefont {Vaulin}},\ }\href
  {https://doi.org/10.1103/PhysRevD.76.124028} {\bibfield  {journal} {\bibinfo
  {journal} {Phys. Rev. D}\ }\textbf {\bibinfo {volume} {76}},\ \bibinfo
  {pages} {124028} (\bibinfo {year} {2007})}\BibitemShut {NoStop}%
\bibitem [{\citenamefont {Sela}(2018)}]{sel18}%
  \BibitemOpen
  \bibfield  {author} {\bibinfo {author} {\bibfnamefont {O.}~\bibnamefont
  {Sela}},\ }\href {https://doi.org/10.1103/PhysRevD.98.024025} {\bibfield
  {journal} {\bibinfo  {journal} {Phys. Rev. D}\ }\textbf {\bibinfo {volume}
  {98}},\ \bibinfo {pages} {024025} (\bibinfo {year} {2018})}\BibitemShut
  {NoStop}%
\bibitem [{\citenamefont {Barcel\'o}\ \emph {et~al.}(2012)\citenamefont
  {Barcel\'o}, \citenamefont {Carballo},\ and\ \citenamefont {Garay}}]{bar12}%
  \BibitemOpen
  \bibfield  {author} {\bibinfo {author} {\bibfnamefont {C.}~\bibnamefont
  {Barcel\'o}}, \bibinfo {author} {\bibfnamefont {R.}~\bibnamefont
  {Carballo}},\ and\ \bibinfo {author} {\bibfnamefont {L.~J.}\ \bibnamefont
  {Garay}},\ }\href {https://doi.org/10.1103/PhysRevD.85.084001} {\bibfield
  {journal} {\bibinfo  {journal} {Phys. Rev. D}\ }\textbf {\bibinfo {volume}
  {85}},\ \bibinfo {pages} {084001} (\bibinfo {year} {2012})}\BibitemShut
  {NoStop}%
\bibitem [{\citenamefont {Polyakov}(1981)}]{pol81}%
  \BibitemOpen
  \bibfield  {author} {\bibinfo {author} {\bibfnamefont {A.}~\bibnamefont
  {Polyakov}},\ }\href
  {https://doi.org/https://doi.org/10.1016/0370-2693(81)90743-7} {\bibfield
  {journal} {\bibinfo  {journal} {Physics Letters B}\ }\textbf {\bibinfo
  {volume} {103}},\ \bibinfo {pages} {207} (\bibinfo {year}
  {1981})}\BibitemShut {NoStop}%
\bibitem [{\citenamefont {Arrechea}\ \emph {et~al.}(2021)\citenamefont
  {Arrechea}, \citenamefont {Barcel{\'{o}}}, \citenamefont {Carballo-Rubio},\
  and\ \citenamefont {Garay}}]{arr21a}%
  \BibitemOpen
  \bibfield  {author} {\bibinfo {author} {\bibfnamefont {J.}~\bibnamefont
  {Arrechea}}, \bibinfo {author} {\bibfnamefont {C.}~\bibnamefont
  {Barcel{\'{o}}}}, \bibinfo {author} {\bibfnamefont {R.}~\bibnamefont
  {Carballo-Rubio}},\ and\ \bibinfo {author} {\bibfnamefont {L.~J.}\
  \bibnamefont {Garay}},\ }\href {https://doi.org/10.1088/1361-6382/abf628}
  {\bibfield  {journal} {\bibinfo  {journal} {Class. Quant. Grav.}\ }\textbf
  {\bibinfo {volume} {38}},\ \bibinfo {pages} {115014} (\bibinfo {year}
  {2021})}\BibitemShut {NoStop}%
\bibitem [{\citenamefont {Fulling}\ \emph {et~al.}(1978)\citenamefont
  {Fulling}, \citenamefont {Sweeny},\ and\ \citenamefont {Wald}}]{ful78}%
  \BibitemOpen
  \bibfield  {author} {\bibinfo {author} {\bibfnamefont {S.~A.}\ \bibnamefont
  {Fulling}}, \bibinfo {author} {\bibfnamefont {M.}~\bibnamefont {Sweeny}},\
  and\ \bibinfo {author} {\bibfnamefont {R.~M.}\ \bibnamefont {Wald}},\ }\href
  {https://doi.org/10.1007/BF01196934} {\bibfield  {journal} {\bibinfo
  {journal} {Communications in Mathematical Physics}\ }\textbf {\bibinfo
  {volume} {63}},\ \bibinfo {pages} {257} (\bibinfo {year} {1978})}\BibitemShut
  {NoStop}%
\bibitem [{\citenamefont {Fabbri}\ and\ \citenamefont
  {Navarro-Salas}(2005)}]{fab05}%
  \BibitemOpen
  \bibfield  {author} {\bibinfo {author} {\bibfnamefont {A.}~\bibnamefont
  {Fabbri}}\ and\ \bibinfo {author} {\bibfnamefont {J.}~\bibnamefont
  {Navarro-Salas}},\ }\href@noop {} {\emph {\bibinfo {title} {{Modeling black
  hole evaporation}}}}\ (\bibinfo  {publisher} {Imperial College Press},\
  \bibinfo {address} {London},\ \bibinfo {year} {2005})\BibitemShut {NoStop}%
\bibitem [{\citenamefont {Levi}\ and\ \citenamefont {Ori}(2015)}]{lev15}%
  \BibitemOpen
  \bibfield  {author} {\bibinfo {author} {\bibfnamefont {A.}~\bibnamefont
  {Levi}}\ and\ \bibinfo {author} {\bibfnamefont {A.}~\bibnamefont {Ori}},\
  }\href {https://doi.org/10.1103/PhysRevD.91.104028} {\bibfield  {journal}
  {\bibinfo  {journal} {Phys. Rev. D}\ }\textbf {\bibinfo {volume} {91}},\
  \bibinfo {pages} {104028} (\bibinfo {year} {2015})}\BibitemShut {NoStop}%
\bibitem [{\citenamefont {Levi}\ and\ \citenamefont {Ori}(2016)}]{lev16b}%
  \BibitemOpen
  \bibfield  {author} {\bibinfo {author} {\bibfnamefont {A.}~\bibnamefont
  {Levi}}\ and\ \bibinfo {author} {\bibfnamefont {A.}~\bibnamefont {Ori}},\
  }\href {https://doi.org/10.1103/PhysRevLett.117.231101} {\bibfield  {journal}
  {\bibinfo  {journal} {Phys. Rev. Lett.}\ }\textbf {\bibinfo {volume} {117}},\
  \bibinfo {pages} {231101} (\bibinfo {year} {2016})}\BibitemShut {NoStop}%
\bibitem [{\citenamefont {Levi}(2017)}]{lev17}%
  \BibitemOpen
  \bibfield  {author} {\bibinfo {author} {\bibfnamefont {A.}~\bibnamefont
  {Levi}},\ }\href {https://doi.org/10.1103/PhysRevD.95.025007} {\bibfield
  {journal} {\bibinfo  {journal} {Phys. Rev. D}\ }\textbf {\bibinfo {volume}
  {95}},\ \bibinfo {pages} {025007} (\bibinfo {year} {2017})}\BibitemShut
  {NoStop}%
\bibitem [{\citenamefont {Crowther}\ and\ \citenamefont {Haro}(2021)}]{cro22}%
  \BibitemOpen
  \bibfield  {author} {\bibinfo {author} {\bibfnamefont {K.}~\bibnamefont
  {Crowther}}\ and\ \bibinfo {author} {\bibfnamefont {S.~D.}\ \bibnamefont
  {Haro}},\ }\href@noop {} {\bibinfo {title} {Four attitudes towards
  singularities in the search for a theory of quantum gravity}} (\bibinfo
  {year} {2021}),\ \Eprint {https://arxiv.org/abs/2112.08531} {arXiv:2112.08531
  [gr-qc]} \BibitemShut {NoStop}%
\end{thebibliography}%

\end{document}